\documentstyle[11pt,amssymb,amsfonts,amsthm,amssymb,amscd,amstext,epsfig,mathrsfs,cite]{article}      
\newcommand{\startappendix}{
\setcounter{section}{0}
\renewcommand{\thesection}{\Alph{section}}}
\newcommand{\Appendix}[1]{
\refstepcounter{section}
\begin{flushleft}
{\large\bf Appendix \thesection: #1}
\end{flushleft}}
\def\ben{\begin{equation}}
\def\een{\end{equation}}

\def\be{\begin{equation}}
\def\ee{\end{equation}}
\def\ba{\begin{array}}
\def\ea{\end{array}}

\def\dalemb#1#2{{\vbox{\hrule height .#2pt
        \hbox{\vrule width.#2pt height#1pt \kern#1pt
                \vrule width.#2pt}
        \hrule height.#2pt}}}

\newcommand{\Dslash}{D\!\!\!\!/\,}
\newcommand{\muRW}{\mu_{\rm RW}}
\newcommand{\TRW}{T_{\rm RW}}
\newcommand{\bea}{\begin{eqnarray}}
\newcommand{\eea}{\end{eqnarray}}

\newcommand{\hm}{\hspace*{-0.6cm}}

\begin{document}

\begin{center}
\vspace{1cm}
{ \LARGE {\bf Holographic Roberge-Weiss Transitions}}

\vspace{1.5cm}

Gert Aarts, S. Prem Kumar and James Rafferty

\vspace{0.8cm}

{\it Department of Physics, Swansea University\\
Swansea, SA2 8PP, UK}

\vspace{0.3cm}
\hspace{1cm} g.aarts, s.p.kumar, pyjames@swansea.ac.uk
\vspace{1cm} \\
May 17, 2010
\vspace{1.5cm}
\end{center}

\begin{abstract}
 We investigate ${\cal N}=4$ SYM coupled to fundamental flavours at 
nonzero imaginary quark chemical potential in the strong coupling and 
large $N$ limit, using gauge/gravity duality applied to the D3-D7 system, 
treating flavours in the probe approximation. The interplay between 
${\mathbb Z}_N$ symmetry and the imaginary chemical potential yields a 
series of first-order Roberge-Weiss transitions. An additional thermal 
transition separates phases where quarks are bound/unbound into mesons. 
This results in a set of Roberge-Weiss endpoints: we establish that these 
are triple points, determine the Roberge-Weiss temperature, give the 
curvature of the phase boundaries and confirm that the theory is analytic 
in $\mu^2$ when $\mu^2\approx 0$. 
 \end{abstract}

\pagebreak
\setcounter{page}{1}
\section{Introduction}
The duality between strongly interacting gauge theories 
and gravity, via the AdS/CFT correspondence \cite{maldacena}, 
has fuelled the exploration of a large number of strongly coupled
systems and their associated thermodynamic phase structures. Essential
features of the thermodynamics of pure gauge theories such as the 
deconfinement transition at strong coupling are captured by
semiclassical gravitational transitions \cite{witten1,
  witten2}. When applied to ${\cal N}=4$ supersymmetric (SUSY)
gauge theory, the gauge/gravity correspondence yields the full phase
diagram at strong coupling, 
as a function of both temperature and chemical potentials for
global R-symmetries \cite{gubser, bcs, Chamblin:1999tk, gc}. Various
qualitative aspects of this phase diagram are also mirrored  by 
the weakly coupled gauge theory \cite{Sundborg:1999ue, Aharony:2003sx,
Yamada:2006rx, Yamada:2007gb, Hollowood:2008gp}. Extensions of  
such holographic computations to include gauge 
theories with matter fields transforming in the fundamental
representation of the gauge group have been considered
\cite{karchkatz, ss} and the resulting phase diagrams have been extensively
studied by various groups \cite{babington, kirsch, ghoroku1, mateos1, albash,
  karch1, mateos2, ghoroku2, mateos3, Nakamura:2007nx, obannon,
  mateos4, Aharony:2006da, Horigome:2006xu}.
\\

In this paper we study the thermodynamics of the $SU(N)$ 
${\cal N}=4$ theory coupled to ${\cal N}=2$ SUSY matter with an {\em
  imaginary} chemical potential for quark (flavour) number, using the  
D3-D7 setup at strong coupling, introduced in \cite{karchkatz}. The
theory has $N_f$  
flavour multiplets preserving ${\cal N}=2$ SUSY, with
$N_f\ll N$, so that flavour loops are suppressed in the large-$N$
theory. The physical motivation and rationale behind the investigation
of thermodynamics with an imaginary chemical potential $\mu_I$ is discussed
below; in short, it provides an additional window into the dynamics
and phase structure of ``QCD-like'' gauge theories with fundamental
matter fields.
\\

\begin{figure}[t]
\begin{center}
\epsfig{file=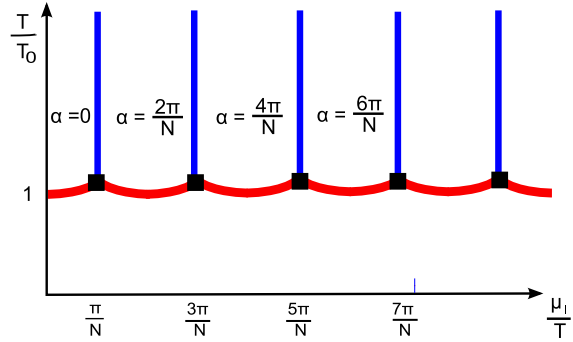, width =3.0in}
\end{center}
\caption{ \small{ 
 Phase diagram in the plane of temperature $T$ and 
imaginary quark chemical potential $\mu_I$. The blue lines represent the 
first order Roberge-Weiss transitions in the high-temperature phase. The 
red line is the line of first-order phase transitions separating the high- 
and the low-temperature phases. The Roberge-Weiss endpoints, indicated by 
the black squares, are triple points at $(\muRW, \TRW)$. The phase of the 
Polyakov loop is denoted with $\alpha$.
 }} 
 \label{phase2} 
 \end{figure}

Our study of the gravity dual description of the
strongly coupled gauge theory reveals the phase diagram depicted in
Figure \ref{phase2}. At high temperatures $T$, there is an infinite
 sequence of first order transitions at $\mu_I=\muRW$, where
\be
\muRW \equiv (2r-1)\frac{\pi T}{N}\,,\qquad r\in{\mathbb Z}\,,
\label{eqmuRW}
\ee
characterized by jumps in the phase of the
 Polyakov loop order parameter. The quarks, {\it i.e.} the fundamental matter
 degrees of freedom, remain unbound and are represented by the
 so-called black-hole embeddings of probe D7-branes in the gravity
 dual setup. 
The sequence of first order transition lines appears due to
 the interplay between ${\mathbb Z}_N$-invariance of the
 adjoint fields and the imaginary quark chemical potential $\mu_I$, 
as was argued on general grounds by Roberge and Weiss in the context
of QCD \cite{rw}. It
 reflects the presence of unbound or deconfined fundamental matter
 degrees of freedom (carrying non-zero $N$-ality). 
A significant feature of the Roberge-Weiss (RW) lines
 is that they must end at some low temperature, below which quarks are
 confined or bound, thus acting as a diagnostic for
 confinement/deconfinement of quarks. The order of the transition
 at the Roberge-Weiss endpoints depends on the 
 theory in question. In the holographic example we study here, the endpoints
 are {\em triple points} since the RW lines meet another line of first 
order
 transitions (indicated in red in Figure \ref{phase2}) at $T=\TRW$. Below 
this line the
 quarks are bound into mesons corresponding to fluctuations of the
 ``Minkowski-embeddings'' of probe D7-branes in the gravity dual
 \cite{kruczenski}. The line of meson-melting transitions is an extension 
to imaginary values of the same phenomenon found in this model for real 
values of the chemical potential \cite{babington, mateos1, mateos2, 
mateos4}. We find that this phase boundary is given by a set of parabola, 
centered at $\mu_I/T= 2\pi r/N$ ($r\in
 {\mathbb Z}$). Explicitly, we find 
 \begin{equation}
 \label{eq:curv}
 \frac{T_c(\mu_I)-T_0}{T_0} = 
\frac{\kappa}{\lambda}\left(\frac{\mu_I}{T_0}\,-\,\frac{2\pi r}{N}\right)^2,
 \;\;\;\;\;\;\;\;\;\;
 \kappa \approx 43.5,
 \end{equation}
when
\begin{equation}
\frac{(2r-1)\pi}{N} < \frac{\mu_I}{T} < \frac{(2r+1)\pi}{N}.
\end{equation}
Here $T_0=T_c(\mu_I=0)$ is the critical temperature for vanishing 
$\mu_I$ and $\lambda=g_{YM}^2N$ is the 't Hooft coupling, which is 
taken to be large. 
The Roberge-Weiss endpoints are therefore located at 
$(\muRW, \TRW)$, where $\muRW$ is given by Eq.(\ref{eqmuRW}) and
 \begin{equation}
 \frac{\TRW}{T_0} = 1 + \frac{\kappa}{\lambda}\frac{\pi^2}{N^2}\,.
\end{equation}
The meson melting temperature at zero chemical potential is determined
by the typical  meson mass scale $M$  in this model and $T_0\approx 0.77M$, where 
$M=2m_q/\sqrt\lambda$ with $m_q$ the
common mass of the ${\cal N}=2$ flavour multiplets \cite{mateos4, kruczenski}.

Since the flavour degrees of freedom are probes in the large-$N$
theory, they do not affect the dynamics of the adjoint degrees of
freedom of ${\cal N}=4$ SYM, which remain deconfined at all
temperatures. Hence, 
the low-temperature phase does not resemble the gluon sector of 
QCD. However, fundamental matter in this system is bound into mesonic 
states at low temperature and unbound at high temperature (with the scale 
set by the meson mass). A notion of a 
low- and a high-temperature phase therefore still exists.
\\

The primary motivation for the exploration of thermodynamics in the
presence of imaginary chemical potentials comes from QCD.
As is well known, the phase structure of QCD at nonzero temperature
and 
quark chemical 
potential is difficult to determine. At large temperature and/or chemical 
potential predictions can be made with perturbation theory, due to 
asymptotic freedom, but most of the phase diagram requires a 
nonperturbative evaluation. Thermal properties ($T>0$) at vanishing 
chemical potential ($\mu=0$) are being investigated vigorously using 
lattice QCD; for a recent review, see {\it e.g.} 
\cite{DeTar:2009ef}. At  
non-zero chemical potential however, a straightforward application of 
lattice QCD runs into problems. Since the fermion determinant is complex, 
algorithms based on importance sampling can no longer be applied, 
resulting in the infamous sign problem. Various approaches to circumvent 
the sign problem in QCD at finite density have been developed over the
past decade; an up-to-date review can be found in  \cite{deForcrand:2010ys}.

The sign problem is due to the nonhermiticity of the Dirac operator. Let 
$\Dslash$ denote the massless Dirac operator, satisfying 
$\gamma_5\Dslash\gamma_5=\Dslash^\dagger$. Integrating out the 
fermion fields then yields the determinant $\det M(\mu) = \det(\Dslash +m 
+\mu\gamma_0)$. Using $\gamma_5$-hermiticity it immediately follows that 
this determinant is complex,
 \begin{equation}
 [\det M(\mu)]^* = \det M(-\mu^*),
 \end{equation}
unless the chemical potential vanishes or is purely imaginary.
 This last observation suggests that lattice QCD simulations at imaginary 
chemical potential $\mu=i\mu_I$ can be performed using standard 
algorithms. Information about the phase structure at real chemical 
potential can then be obtained by analytic continuation: since the 
partition function is an even function of $\mu$, analytical continuation 
from imaginary $\mu$ ($\mu^2<0$) to real $\mu$ ($\mu^2>0$) is in principle 
straightforward. This approach has indeed been used extensively 
\cite{deForcrand:2002ci,deForcrand:2003hx,deForcrand:2006pv, 
deForcrand:2010he,D'Elia:2002gd,D'Elia:2004at,D'Elia:2007ke, 
D'Elia:2009tm,D'Elia:2009qz,Cea:2010md}.

 As mentioned above, 
the phase structure at nonzero imaginary chemical potential is nontrivial 
due to the interplay between the ${\mathbb Z}_N$ symmetry and $\mu_I$ \cite{rw}. 
As we will review below, the partition function is periodic in 
$\mu_I$ and satisfies
 \begin{equation}
 Z[\mu_I] = Z\left[\mu_I+\frac{2\pi r T}{N} \right], 
\;\;\;\;\;\;\;\;r\in \mathbb{Z}.
 \end{equation}
Combined with reflection symmetry $Z[\mu_I]=Z[-\mu_I]$, this leads to 
Roberge-Weiss transitions at $\mu_I=\muRW$  (Eq.(\ref{eqmuRW})). 
 At these values of $\mu_I$, the 
theory undergoes a transition between adjacent ${\mathbb Z}_N$
sectors, which can be distinguished by the phase of the Polyakov loop.

At high temperature the Roberge-Weiss transitions are first order, 
resulting in nonanalytic behaviour of observables at $\mu_I=\muRW$.
At low temperature, the Roberge-Weiss periodicity 
is smoothly realized and no nonanalyticity is present. This was first 
suggested in \cite{rw} and has been confirmed with 
lattice QCD simulations 
\cite{deForcrand:2002ci,deForcrand:2003hx,deForcrand:2006pv, 
deForcrand:2010he,D'Elia:2002gd,D'Elia:2004at,D'Elia:2007ke, 
D'Elia:2009tm,D'Elia:2009qz,Cea:2010md}. The Roberge-Weiss lines 
therefore end.
 In QCD the high- and low-temperature phase are separated by a thermal 
confinement/deconfinement transition,\footnote{Although strictly 
speaking it is only a confinement/deconfinement transition when the quarks 
are infinitely heavy, we use this terminology for light 
quarks as well.}
 which extends into the $\mu_I$ direction. It is widely assumed that this 
transition line will join the Roberge-Weiss lines at $\mu_I=\muRW$. 
Generically the phase diagram will therefore be as depicted in Fig.\ 
\ref{phase2}.

In QCD properties of the confinement/deconfinement transition depend
crucially on the number of flavours and the quark masses. For instance,
for three degenerate flavours, the transition is first order for light and
heavy quarks (which is tied to the chiral and centre symmetries
respectively), while for intermediate quark masses, the transition is a
crossover. It has only recently been demonstrated that the Roberge-Weiss
endpoint is a triple point for two \cite{D'Elia:2009qz} and three
\cite{deForcrand:2010he} degenerate light or heavy flavours, while it is a
second order endpoint in the 3d Ising universality class for intermediate
quark masses. Moreover, in a beautiful paper \cite{deForcrand:2010he}, de
Forcrand and Philipsen argued in the case of heavy quarks that the
tricritical point found in the quark mass-temperature plane at
$\mu_I=\muRW$ dictates the deconfinement critical line for real $\mu$ by
tricritical scaling.

It is therefore clear that an understanding of the phase diagram at 
imaginary chemical potential can provide important insight into the phase 
structure at real chemical potential. For this reason this topic has been 
under intense investigation, using both lattice QCD as mentioned above, renormalization-group studies \cite{Braun:2009gm}, as 
well as effective models 
\cite{Bluhm:2007cp,Sakai:2008um,Kashiwa:2008bq,Kouno:2009bm}.
In this paper we take a third approach and apply 
gauge-gravity duality to study the phase diagram of a strongly coupled 
gauge theory at imaginary chemical potential.

This paper is organized as follows. We begin by reviewing the
arguments of \cite{rw} in Section 2 and summarize results that can be
obtained at arbitrarily weak gauge coupling. In Section 3 we review
the well known properties of the gauge theory model we consider (${\cal N} =4$ SYM
coupled to 
${\cal N}=2$ matter). Section 4 contains the holographic description of the
model at imaginary chemical potential and a numerical study of the
configurations that dominate the ensemble as a function of temperature
and $\mu_I$. The resulting phase diagram is constructed in Section 5.
Section 6 contains a brief outlook. In the Appendix we outline, in some detail, the
thermodynamics of the model at weak coupling.

\section{Roberge-Weiss phase transitions}

It is well known that the deconfinement phase transition in $SU(N)$
gauge theories without   
matter in the fundamental representation can be precisely
characterized, in the Euclidean formulation, by the expectation value of the 
thermal Wilson line or  Polyakov loop  $P=\frac{1}{N}{\rm Tr}
\exp(i\oint A_\tau )$. The ${\mathbb Z}_N$ symmetry of the pure gauge
theory under which $P\to P e^{2\pi i k/N}$, ($k=0,1,2,\ldots N-1$) is
spontaneously broken by an expectation value for the Polyakov loop in
the high temperature deconfined phase. This distinction between
confining and deconfined phases is lost upon the
introduction of matter in the fundamental representation of the gauge
group since there is no ${\mathbb Z}_N$ symmetry and the operator $P$ can
obtain an expectation value at all temperatures. However, as was
argued in \cite{rw}, there is a remnant of this symmetry which can
still be used to distinguish between the phases in which 
quarks are confined or deconfined.

The idea of \cite{rw} is that in theories with fundamental
matter and an appropriately defined baryon number $U(1)_B$ symmetry, 
the difference
between the confined and deconfined phases can manifest itself in the
behaviour of the theory as a function of an {\em imaginary} chemical potential
for baryon or quark number. 
With an imaginary chemical potential $\mu_I$ for quark number, the
thermal partition function  
\be
Z[\mu_I] = {\rm Tr} \left(e^{-\beta H + i \beta \mu_I  N_q}\right)
\ee
is naturally periodic under $\mu_I\to \mu_I +2\pi T$ $(\beta\equiv 1/T)$, 
since $N_q$ is quantized.  Hence $\frac{\mu_I}{T}\in [-\pi,\pi]$.

There is a further periodicity of this partition function following from the
${\mathbb Z}_N$ symmetry of the adjoint degrees of freedom \cite{rw} and
it can be understood as follows. We recall that the (imaginary)
chemical potential for fermion number couples to
fundamental matter fields in the same way as the time component $A_\tau$
of the $SU(N)$ gauge field. The imaginary chemical potential $\mu_I$
is introduced by the replacement $\partial_\tau+ iA_\tau \,\to\,
\partial_\tau+i A_\tau - i \mu_I\,$, acting on matter fields in the
fundamental representation.

This $\mu_I$-dependence at the level of the action can be removed by a
phase redefinition of fields $\phi$ charged under $U(1)_B$, and
transferred into boundary conditions around the thermal circle,
\be
\phi(\vec x,\beta)=\pm e^{i\mu_I/T}\,\phi(\vec x,0 ),
\ee
where the $+$ and $-$ signs apply to bosonic and fermionic modes respectively.
We can perform a 
second variable change, corresponding to a gauge transformation by
an element $U(\vec x, \tau)$ of $SU(N)$, with the property $U(\vec
x,\beta)= e^{2\pi i r/N}\,U(\vec x, 0)$, $r\in {\mathbb Z}$. This 
leaves the action
and the path integral measure invariant, whilst only altering the
boundary conditions  
for the fields around the thermal circle,
\be
\phi(\vec x,\beta) = \pm e^{i\mu_I/T}\,e^{i2\pi r/N}\, \phi(\vec x, 0).
\ee
Since the partition function is left invariant by variable changes, we
conclude that
\be
Z[\mu_I] = Z\left[\mu_I + \frac{2\pi r T}{N} \right]\,,\qquad
r \in {\mathbb Z}.
\ee
This implies that $\frac{\mu_I}{T} \in
[-\frac{\pi}{N},\frac{\pi}{N}]$. 

At high temperatures, when the weak coupling approximation is consistent, 
a perturbative evaluation of the free energy 
$F[\mu_I] = - \ln[Z]/\beta$ shows 
first-order phase transitions as a function of $\mu_I$ at  $\mu_I=
\muRW= (2r-1)\pi T/N$ ($r\in{\mathbb Z}$). These are the Roberge-Weiss
transitions.  At low temperatures, when the theories are typically
strongly coupled, lattice studies suggest that $F[\mu_I]$ is a smooth
function of $\mu_I/T$. The situation at very high and very low
temperatures is therefore expected to be qualitatively as in Figure
\ref{phase2}. The nature of the transition between these two behaviours depends on
the detailed dynamics of the theories in question, as we have already 
discussed in the Introduction.

\subsection{Roberge-Weiss transitions at weak coupling}
At suitably high temperatures when the gauge theory is in a deconfined
phase it is generically possible to compute the effective potential as
a function of $\mu_I$ perturbatively \cite{weiss}.  The details of
this calculation at the one-loop level are summarized in the
Appendix. Here we state the result of this calculation. The basic idea
involves computing an effective potential for the eigenvalues of the
holonomy around the thermal circle (the Polyakov loop matrix), $\exp
(i\oint A_\tau)$ in the presence of fundamental matter. The effective
potential at one-loop is the sum of a gluonic (adjoint matter) piece
and a flavour (fundamental matter) piece. While the gluonic
contribution, $V_A$, is ${\mathbb Z}_N$ invariant and independent of $\mu_I$,
the terms arising from the flavour fields, $V_f$, are not ${\mathbb Z}_N$
invariant and explicitly depend on the imaginary chemical potential,
\be
V_{\rm eff}(\,\{\alpha_i\}\,) = V_A\left(\,\{\alpha_{ij}\}\,\right) + V_f(\,\{\alpha_i -\mu_I/T\}\,)\,\,,\qquad \alpha_{ij}\equiv\alpha_i-\alpha_j.
\ee
Here $e^{i \alpha_j}$, ${j=1,2,\ldots N}$ are the eigenvalues of the
Polyakov loop matrix. Generically, at high temperatures, in the
absence of fields in the fundamental of $SU(N)$, $V_A$ will
have $N$ minima, each of which breaks the ${\mathbb Z}_N$ symmetry
spontaneously. At these minima $\alpha_i= 2\pi k/N$ for
$k=0,1,2,\ldots N-1$. Introduction of fundamental matter turns these
into local minima with a single global minimum at $\alpha_i=0$ for
vanishing $\mu_I$. Since $V_f$ depends only on the combination $(\alpha_i-\mu_I/T)$, 
 the global minimum will move to  $\alpha_i=2\pi r/N$ when
$\mu_I/T = 2\pi r /N$ ($r\in {\mathbb Z}$).
 We conclude that the different local minima of
the effective potential compete as $\mu_I$ is varied
and there are first order phase transitions between two
neighbouring minima whenever $\mu_I/T = (2r-1)\pi/N$ (${r\in
{\mathbb Z}}$). The perturbative result for the free energy across these
transitions is computed in the Appendix. We quote the result here,
specifically for a large-$N$ gauge theory coupled to both adjoint and
fundamental matter fields. For such a theory with $n_f$ complex
scalars and $\tilde n_f$ Weyl
fermions, all in the fundamental representation of $SU(N)$ and
with mass $m_q$, 
we find the free energy as a function of $\mu_I$ to be
\be
F[\mu_I]= N\, \frac{T^4}{12} \,\left[n_f \,f_{\rm
    bose}\left(\frac{m_q}{T}\right)  \,+\, 
\tilde n_f \,f_{\rm fermi}\left(\frac{m_q}{T}\right) \right]\,
\min_{r\in {\mathbb Z}}\left(\frac{\mu_I}{T}-\frac{2\pi r}{N}\right)^2,
\label{fweak}
\ee
where
\be
f_{\rm bose}\left(x\right)= \frac{3}{2\pi^2}\int_0^\infty y^2 \,dy\,
\frac{1}{\sinh^{2}\left(\frac{1}{2}\sqrt{y^2+x^2}\right)\,,}
\label{fb}
\ee
and
\be
f_{\rm fermi}
\left(x\right)= \frac{3}{2\pi^2}\int_0^\infty y^2 \,dy\,
\frac{1}{\cosh^{2}\left(\frac{1}{2}\sqrt{y^2+x^2}\right)}\,.
\label{ff}
\ee
These functions are determined by the Bose-Einstein and
Fermi-Dirac distributions for free particles. 
As $m_q/T \to 0$, $f_{\rm bose}$ and $f_{\rm fermi}$ 
approach $2$ and $1$
respectively. 

\begin{figure}[h]
\begin{center}
\epsfig{file=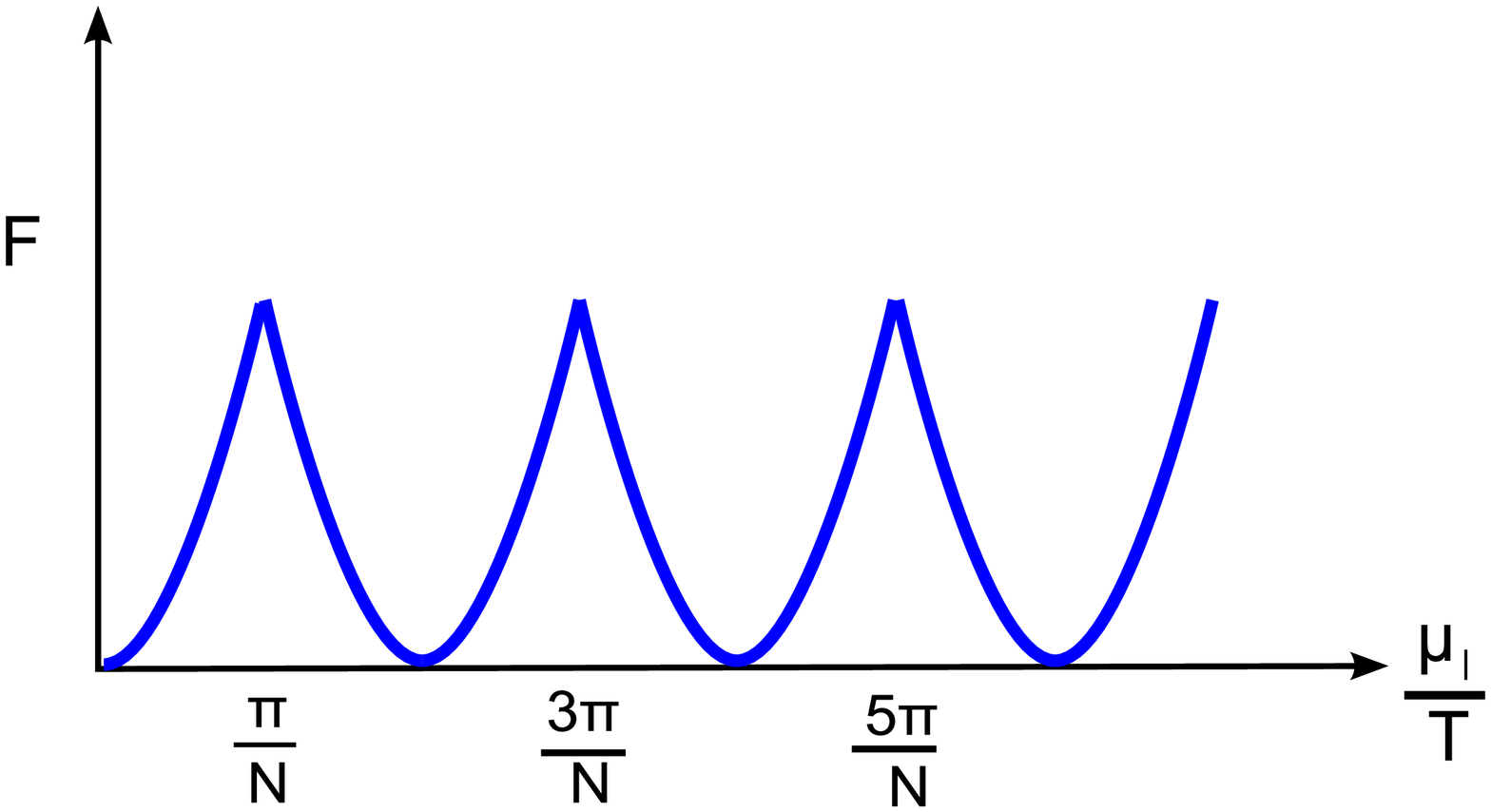, width =2in}
\hspace{0.5in}
\epsfig{file=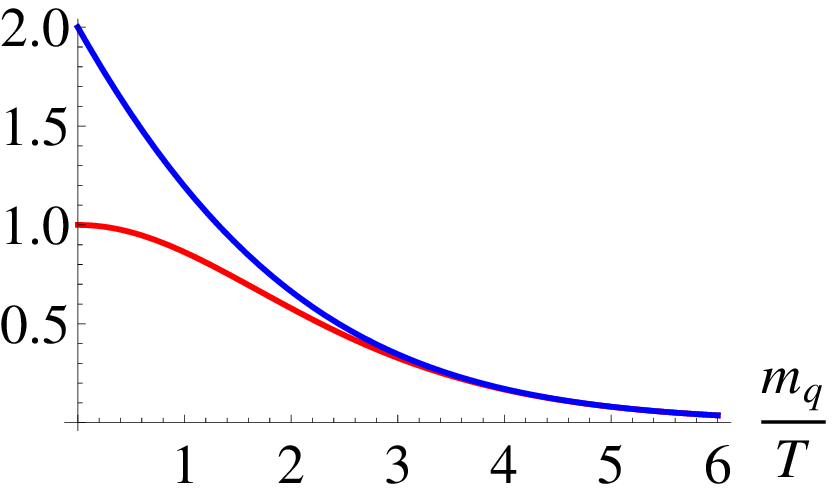, width=2in}
\end{center}
\caption{ \small{\underline{Left}: The generic behaviour of the free
    energy across Roberge-Weiss transitions. \underline{Right:} The functions
    $f_{\rm bose}$ (blue) and $f_{\rm fermi}$ (red) versus the ratio $m_q/T$.}} 
\label{cusp}
\end{figure}

The free energy is a function with cusps (Figure \ref{cusp}) located at odd
integer multiples of $\pi/N$ where phase transitions occur. It is a
quadratic function only because we chose to look at 
the large-$N$ limit which forces
the Roberge-Weiss transition points to be closely spaced
within intervals of $2\pi/N$. Hence only the second derivative
of the free energy  
function at its minimum is relevant. The strength of the transition,
as measured by the coefficient functions $f_{\rm bose}$ and 
$f_{\rm fermi}$ varies with 
the ratio $m_q/T$. We have plotted these in Figure
\ref{cusp}. Notable features are their behaviours for small and large
masses. $f_{\rm bose}$ is determined by the Bose-Einstein distribution which
diverges for small masses and is responsible for the linear
dependence on $m_q/T$ when $m_q\ll T$. For
large masses (or low temperatures), both coefficients are
exponentially Boltzmann-suppressed.

We note here that at weak coupling 
only the ratio $m_q/T$ is relevant. Typically however,
nonperturbative physics will generate another scale 
(such as $\Lambda_{\rm QCD}$ in QCD). In that case the
notion of high and low temperatures and masses has
an independent meaning and the weak coupling result can
be trusted when $T \gg \Lambda_{\rm QCD}$.
 In the model we study in this paper, 
the 't Hooft coupling $\lambda=g^2_{YM} N$ can be dialled to
any fixed value (assuming the number of flavours is small compared to $N$),
and there is no dynamical scale. This means that the ratio $m_q/T$ is
the only  parameter which can be varied at fixed 't Hooft coupling and
we expect the above picture to be valid for $\lambda\ll 1$. For large
$\lambda$, the perturbative evaluation is clearly not valid and we
must use the dual gravity picture.

\section{${\cal N}=4$ theory coupled to ${\cal N}=2$ flavours}
We will consider the ${\cal N}=4$ theory in flat space, and at finite
temperature ,  
coupled to ${\cal N}=2$ supersymmetric matter
transforming in the fundamental representation of the $SU(N)$ gauge
group. When the number, $N_f$, of ${\cal N}=2$ hypermultiplets is
small ($N_f\ll N$), the effect of these hypermultiplets on the dynamics
of the ${\cal N}=4$ matter (such as the running of the gauge coupling) can
be consistently neglected and the matter fields can be treated in a
``probe'' approximation. The ${\cal N}=4$ degrees of freedom are
always in the deconfined phase at finite temperature in flat space. 
This particular theory can be readily studied at
strong coupling using the AdS/CFT correspondence as was first pointed out in
\cite{karchkatz}.

\subsection{The classical theory}
The ${\cal N}=4$ theory can be naturally coupled to matter in the
fundamental representation and thereby preserve a maximum of eight
supercharges  (${\cal N}=2$ SUSY). In the language of ${\cal
  N}=1$ superfields, the ${\cal N}=2$ hypermultiplet consists of two
chiral superfields $(Q,\tilde Q)$ transforming in the $(\bar N, N)$
representation of the $SU(N)$ gauge group. The superfields in the
hypermultiplet each consist of a squark and a quark $(q,\psi)$ and
$(\tilde q, \tilde\psi)$.

Introducing $N_f$
hypermultiplets $(Q^i,\tilde Q_i)$ and using the ${\cal N}=1$ SUSY notation,
the superpotential of the theory with flavours can be written as
\be
W= {1\over g^2_{YM}} \left(\sum_{i=1}^{N_f}\left(\sqrt2 \,\tilde
    Q_i\,\Phi_3\, Q^i + m_q \,\tilde Q_i\,Q^i\right) + \sqrt 2 \,{\rm
    Tr}(\Phi_3\,[\Phi_1,\Phi_2])\right). 
\ee
The three adjoint chiral multiplets of the ${\cal N}=4$ theory are
$\Phi_1, \Phi_2$ and $\Phi_3$. 
After coupling to the ${\cal N}=2$ hypermultiplets,  
$\Phi_3$ naturally becomes the scalar part of the ${\cal N}=2$ vector
multiplet, while $\Phi_{1,2}$ together constitute an ${\cal N}=2$
adjoint hypermultiplet.  
Note that while it is possible to introduce distinct masses for
each of the $N_f$ flavours in question, we will only consider the case
where all masses are equal. For equal quark masses the theory has a
$U(N_f)\simeq U(1)_B\times SU(N_f)$ flavour symmetry group. The
$U(1)_B$ is identified as the ``baryon number'' symmetry\footnote{For 
massless quarks the theory has the global symmetry group,
$U(1)_B\times SU(N_f)\times 
SU(2)_R \times U(1)_R\times SU(2)_{\Phi}.$
The $SU(2)_R\times U(1)_R$ factor is the R-symmetry group of the ${\cal N}=2$
gauge theory.} under which $Q^i$ and $\tilde Q_i$ transform with equal and
opposite charges. 
At the level of the Lagrangian, the mass term for the
hypermultiplet induces the operator,
\be
{\cal L} \supset \frac{1}{g^2_{YM}}\left(
m_q\,{\tilde \psi}^i\psi_i + m_q^2 q_i^\dagger q^i +m_q^2 \tilde q_i^\dagger \tilde q^i +\sqrt 2 m_q q_i^\dagger\Phi_3 q^i +
+ \sqrt 2 m_q \tilde q_i \Phi_3 \tilde q^{i\dagger}+ \,{\rm h.c.}
\right).
\ee
In what follows, the expectation value of the 
operator $\Sigma_m \equiv \frac{\partial {\cal L}}{\partial m_q}$ 
will be referred to as the  ``quark condensate''.

\subsection{D-brane picture}
\label{secD}
 At weak coupling (and zero temperature) 
in the D-brane picture in type IIB string theory, the ${\cal N}=2$ matter
hypermultiplets can be introduced by placing $N_f$ coincident D7-branes in the
presence of $N$ coincident D3-branes. The low energy dynamics of the
latter is ${\cal N}=4$ SUSY gauge theory. The D3 and D7-brane
worldvolumes span the coordinates indicated below in the ambient ten
dimensional spacetime,

\begin{center}
\begin{tabular}{ccccccccccc}
Coordinates & $x_0$ & $x_1$ & $x_2$ & $x_3$ & $x_4$ & $x_5$ & $x_6$ & $x_7$ & $x_8$ & $x_9$ \\
D3 & $\times$ &  $\times$  & $\times$ & $\times$ \\ 
D7 & $\times$ &  $\times$  & $\times$ & $\times$ & $\times$ &  $\times$  & $\times$ & $\times$
\end{tabular}
\end{center}

The fluctuations of the open strings stretching between the
D3  and D7-branes yield the $N_f$ hypermultiplets transforming in the
fundamental representation of the $SU(N)$ gauge symmetry of the 
D3-branes. These also transform under the 
$U(N_f)\simeq U(1)_B\times SU(N_f)$ symmetry group associated to the
D7's, which appears as a flavour symmetry at low energies.
 The separation in the $x_8$-$x_9$ plane, 
between the stack of D3-branes and the flavour branes
translates to a mass $m_q$ for the hypermultiplets. For a separation
$\ell$ in this plane, the hypermultiplet mass is given by the string
tension times the separation, $m_q=\frac{\ell}{2\pi\alpha'}$.

\section{Thermodynamics at strong coupling }
In the large-$N$ and strong 't Hooft coupling limit
($\lambda=g^2_{YM}N\gg 1$), the fundamental hypermultiplets
are introduced as probe D7-branes into the geometry sourced by the $N$
D3-branes, namely $AdS_5\times S^5$ \cite{karchkatz}. 
At finite temperature, in the
deconfined phase, the corresponding geometry is the Schwarzschild black hole in 
$AdS_5 \times S^5$ (with flat horizon). Wick rotating to Euclidean
signature, the spacetime metric is
\be
ds^2 = \left( f(r)d\tau^2+\frac{dr^2}{f(r)}+ \frac{r^2}{R^2} 
d{\vec x}^2 + R^2 \,d\Omega_5^2\right),
\ee
where
\be
f(r)=\frac{1}{R^2}\left(r^2-\frac{ r_H^4}{r^2}\right)\,,\qquad
r_H=\pi T R^2,\qquad R^4 = 4\pi(g_s N)\alpha^{\prime 2}=
g^2_{YM}N\,\alpha^{\prime 2}\,,  
\ee
and $R$ is the AdS radius.

In the Euclidean picture the geometry ends
smoothly at $r=r_H$, provided we make the periodic thermal identification 
$\tau \simeq \tau + 1/T$. Topologically, the Euclidean black hole spacetime is 
${\mathbb R}^3\times D_2 \times S^5$.

As suggested by the weak coupling setup in Section (\ref{secD}), the
flavour  D7-branes must wrap an $S^3\subset S^5$ while filling the $AdS_5$
directions \cite{karchkatz, babington}.  
The embedding of the D7-brane is characterized by the
``slipping angle'' which determines the size of the three-sphere
wrapped by the D7-brane. The slipping angle $\theta$ is defined by
the following paramerization of the $S^5$ metric
\begin{equation}
d\Omega_5^2=
d\theta^2+\sin^2\theta\,d\phi^2+\cos^2\theta\,d\Omega_3^2\,;\qquad 
0\leq\theta\leq\frac{\pi}{2}\,.
\label{s5}
\end{equation}
It is more convenient to rewrite the background metric in a
Fefferman-Graham type coordinate system \cite{FG, karchskenderis}. 
We use the conventions of
\cite{mateos2}, to write the black hole metric in the form
\begin{eqnarray}
&&ds^2=\frac{1}{2}\,\frac{\rho^2}{R^2}\,\left(\,\frac{h^2}{\tilde h} \,d\tau^2
+\tilde h \,{d\vec x}^2\right)+\frac{R^2}{\rho^2}\,(d\rho^2+\rho^2
\,d\Omega_5^2)\,,\\\nonumber\\\nonumber
&& h(\rho)\,\equiv\,1-\frac{r_H^4}{\rho^4}\,,\qquad \tilde
h(\rho)\,\equiv\,1+\frac{r_H^4}{\rho^4}.
\end{eqnarray}
This form of the metric is obtained after the coordinate change 
$r^2=\frac{1}{2}\left(\rho^2+r_H^4/\rho^2\right)$, and we take the
metric of the five-sphere as in Eq.(\ref{s5}). The slipping angle
$\theta$ will be dual to the quark mass operator $\Sigma_m$, and
depend only on the radial coordinate of AdS space,
\be
\theta=\theta(\rho).
\ee
This parametrization is convenient for making 
contact with the weak coupling D-brane picture. Focussing on the 
six directions transverse to the D3-branes, we write
\be
d\rho^2+\rho^2 d\Omega_5^2\,= \,d\rho^2_2+ \rho_2^2 \,d\phi^2+
d\rho_1^2+\rho_1^2\,d\Omega_3^2\,,
\label{sixd}
\ee
with
\be
\rho_1=\rho\cos\theta\,,\qquad\rho_2=\rho\sin\theta.
\ee
The $x_8-x_9$ plane can now be identified with the $(\rho_2,\phi)$
plane. We choose the D7-brane to be at $\phi=0$, {\it i.e.} the $x_8$
axis. The asymptotic value of the D3-D7 separation, given by $\rho_2$ 
for large $\rho$, yields the hypermultiplet mass $m_q$. 
 
\subsection{${\mathbb Z}_N$ breaking in thermal ${\cal N}=4$ theory}
As argued in \cite{aharonywitten}, one may understand the spontaneous
breaking of the ${\mathbb Z}_N$ center symmetry at high temperatures
as follows . We first note that the disk $D_2$ in the black hole geometry
naturally allows  turning on a
Neveu-Schwarz potential $B$ with $dB =0 $ and
\be
\alpha \equiv\frac{1}{2\pi l_s^2} \int_{D_2} \, B.
\ee
This yields a phase $e^{i\alpha}$ for the Polyakov-Maldacena
loop\footnote{Strictly speaking, the Wilson loop computed by the
  wrapped 
  string world-sheet is the supersymmetric version of the standard
  Polyakov loop and includes the adjoint scalar fields of ${\cal N}=4$
SYM in its definition, ${\cal U} \equiv \exp i\oint (A_\tau - i \theta^I(\tau)
\Phi^I)$. This is an important distinction in principle, but will not
influence our discussion below, as the Polyakov-Maldacena loop has the
same transformation properties under ${\mathbb Z}_N$ transformations
as the standard Polyakov loop order parameter.}
\cite{malpol}
computed by wrapping a
fundamental string worldsheet on the cigar $D_2$. Thus $\alpha$ is defined modulo $2\pi$. 
One then finds that
an effective potential for $\alpha$ is induced if we consider terms in
the type IIB effective action that depend on the RR 3-form $F_{(3)}$,
\be
S_{\rm IIB} \supset\frac{1}{(2\pi)^7 l_s^{8}}\left(\int d^{10}x\,\sqrt{g}\,
\frac{1}{12} F_{(3)}^2 - \int F_{(5)}\wedge F_{(3)}\wedge B\right ).
\ee
Integrating over $S^5 \times D_2$, we may obtain the effective theory
for the $F_{(3)}$ field on  ${\mathbb R}^3$. Since $F_{(3)}$ has only
one non-zero component $F_{123}$ on ${\mathbb R}^3$ we have
\be
{S}_{\rm eff}[F_{(3)}]= 
\int
\,d^3x\,\left[\frac{\lambda}{T^3}\frac{1}{2^8\pi^6\,l_s^4}
\left(F_{123}\right)^2 - \frac{N\alpha}{4\pi^2 l_s^2} F_{123}\right],
\ee
where we have used the fact that the type IIB background has $N$ units
of five-form flux through $S^5$, $\int_{S^5} F_5/(2\pi
l_s)^4 = N$. Adapting the arguments arising in the context of $U(1)$
gauge theory in two dimensions to the present situation \cite{aharonywitten}, gauge
invariance of quantum states implies the quantization of
$F_{123}$. States are characterized by a constant (quantized) value
for the field strength (similarly to constant electric fields in two
dimensions) and one finally  finds the effective potential for the
phase of Polyakov-Maldacena loop to be
\be
 V_A(\alpha) =  4\pi^2 N^2 T^4\,
\frac{1}{\lambda}\,\min_{r\in {\mathbb Z}} \,
\left(\alpha-\frac{2\pi}{N} r\right)^2. 
\label{znpot}
\ee 
The potential is minimized when $\alpha=2\pi r/N$ for $r\in {\mathbb
  Z}$ and this leads to a spontaneous breaking of the ${\mathbb Z}_N$
symmetry of the ${\cal N}=4$ theory in the deconfined, high
temperature phase. Note that this effective potential is
${\cal O}(N^2)$, as expected for a leading order large-$N$ effect and
proportional to $T^4$, as expected from dimensional
analysis. Interestingly, it is also proportional to $1/\lambda$ in the
strongly coupled theory.
\subsection{The effect of flavours}
Now we would like to calculate how the effective potential for
$\alpha$ is modified upon introducing fundamental flavours and in the
presence of an {\it imaginary} chemical potential for quark 
number. The basic effect of introducing flavour D7-branes was already addressed
in \cite{yee} and was argued to lift the degeneracy of the $N$ vacua
discussed above.  The overall dynamics of $N_f$ flavour D7-branes,
ignoring the non-Abelian world-volume degrees of freedom in the
probe limit ($N_f\ll N$), can be described by the (Euclidean)
Dirac-Born-Infeld action 
\be
S_{D7}= N_f\, T_{D7}\int d^8\xi e^{-\Phi}\,\sqrt{{\rm det}({}^*g +
  (2\pi\alpha'F + {}^*B))}\,,
\ee
where ${}^*g$ and ${}^*B$ are the pullbacks of the spacetime metric
and the Neveu-Schwarz two-form potential on to the D7-brane
worldvolume. In addition we have that the dilaton $e^{-\Phi}=1$ in the background 
dual to the ${\cal N}=4$ SUSY gauge theory. The D7-brane tension can
be written in terms of gauge theory parameters $\lambda$, $N$ 
(and $R$, the AdS radius) as
\be
T_{D7}= \frac{1}{\,g_s \alpha'^4 (2\pi)^7} = \frac{\lambda N}{32\pi^6 R^8 }.
\ee

We will take the D7-brane probe to wrap the $S^3$ in Eq.(\ref{s5}) at
an angle $\theta$ and fill the directions $\tau, \vec x$
and $\rho$. For the most part, we will primarily be interested in classical
configurations that completely wrap $D_2$, the cigar in the Euclidean black
hole,  corresponding to ``melted meson'' states \cite{mateos1, mateos2}. 
For these configurations it will be consistent to turn on a
field strength $F$ along the world-volume. A real value for $F$ in the
Euclidean action will
correspond to an imaginary gauge potential upon Wick rotating back to
Lorentzian signature.

\subsection{Imaginary chemical potential}
In the boundary field theory formulated on Minkowski space, a chemical
potential for a global $U(1)$ symmetry can be thought of as follows: 
We imagine that the global symmetry is gauged and then turn  
on a constant background value for the time component of the gauge
field, while setting other components to zero. A real chemical potential 
is obtained by assigning an imaginary constant value for this
background gauge field in the Euclidean formulation.
Equivalently, to get {\em imaginary} chemical
potential for the $U(1)_B$ global symmetry, in the Euclidean theory
we must introduce a real value for the time component of the 
background gauge field associated to $U(1)_B$. This is very natural in
the AdS/CFT framework.

The gauge field associated to the 
overall $U(1)$ factor describing the center-of-mass dynamics of
the probe D7-branes, is dual to $U(1)_B$. In the Euclidean theory, the
imaginary chemical potential $\mu_I$ corresponds to a constant
real boundary value for the gauge field $A_\tau$ on the world-volume of the
D7-brane,
\be
\mu_I= - T \int_0^\beta\,d\tau \int_{r_H}^\infty
d\rho\,F_{\tau\rho}=\lim_{\rho\to\infty} A_{\tau}(\rho). 
\ee
Here we choose the gauge freedom to set $A_\rho=0$ and we 
require that $A_\tau(r_H)=0$ in order to avoid a singularity
at the tip of the cigar. We are implicitly assuming that the
D7-brane wraps the black hole cigar $D_2$ completely. The case where
it turns back before getting to the horizon will be discussed
separately. 

The action for the D7-brane probe configuration described above is
\be
S_{D7}=2\pi ^2 N_f T_{D7}\,\int \beta\,d^3 x \int_{r_H}^\infty
d\rho\left(\frac{\rho^3}{4}\,h\,\tilde h\,\cos^3\theta 
\sqrt{1+ \rho^2\theta'(\rho)^2 + 2(2\pi \alpha'F_{\tau\rho}
  +{}^*B_{\tau\rho})^2\frac{\tilde h}{h^2}}\right). 
\label{dbiansatz}
\ee
We note that in the absence of a $B$ field in the background, this is
the same action used in \cite{mateos2}, provided we make the
replacement $F\to i F$. The qualitative nature of the family of
solutions is, however, different to that encountered in
\cite{mateos2} for real baryon number chemical potential.

\subsubsection{Periodicity in $\mu_I$}

The periodicity in $\mu_I$ discussed earlier is naturally built
into the supergravity dual because the D-brane action only depends on the
combination $B+2\pi \alpha^\prime F$. This means that the classical
solutions to the DBI equations of motion will be charactarized by the
combination 
\be
\alpha -\frac{\mu_I}{T} = \int_{D_2} \left(F+ \frac{B}{2\pi 
\alpha'}\right).  
\ee
A shift of $\mu_I$ by $2\pi T$ can be absorbed by $\alpha \to \alpha
+2\pi$. Hence $\mu_I$ is only defined modulo $2\pi T$ as expected from
general considerations. 

As at weak coupling, there is a further periodicity visible in the
finite temperature theory at strong coupling.
The shift $\mu_I \to \mu_I + 2\pi T/N$ can be compensated by a corresponding
shift of $\alpha$ by the same amount, leaving the physics unchanged
since the ``glue'' sector of the theory (${\cal N} = 4$ matter) is
invariant under such shifts of $\alpha$ (from Eq.(\ref{znpot})). Note that
this ``periodicity'' involves a jump from one ${\mathbb Z}_N$ vacuum
to another.
Therefore,
this system should be expected to display the infinite sequence of
Roberge-Weiss transitions as the effective potential will be a sum
of two pieces, one of which is ${\mathbb Z}_N$ invariant, 
while the other piece, induced by the fundamental flavours, depends
only on the combination $\alpha-\mu_I/T$.

\subsection{Solving the DBI equations of motion}
The solutions of the D3/D7 system at finite temperature, with and 
  without a {\em real} chemical potential, have been extensively
studied in earlier works \cite{babington, kirsch, ghoroku1, mateos1, albash,
  karch1, mateos2, ghoroku2, mateos3, obannon, andythesis}. 
Our aim, however, is to study
the solutions arising in the presence of the imaginary chemical potential 
$\mu_I\neq 0$. These will be distinct from the ones found, for
example, in \cite{mateos2,ghoroku2,obannon}. Specifically, for a real
chemical potential, the  term under the square root in
Eq.(\ref{dbiansatz}) would be of the form
$\sqrt{1+\rho^2\theta^{\prime 2}- 2(2\pi\alpha'F)^2\tilde h
  /h^2}$, yielding solutions with different behaviour to the ones
we will find below.

From the solutions to the DBI equations of motion following from
Eq.(\ref{dbiansatz}), we 
will reconstruct the dependence of the effective potential on $\alpha-\mu_I/T$, for generic quark masses.

We work in the gauge $A_\rho=0$, so that
$F_{\tau\rho}=-\partial_\rho A_\tau$. 
Since the action (\ref{dbiansatz}) does not depend explicitly on
$A_\tau$, the conjugate momentum is conserved. In the usual situation
with real chemical potentials, the variable conjugate to $A_\tau$ is
the quark density. We will continue to refer to this conjugate
variable as the ``charge density''  $d$, in the situation with 
an imaginary chemical potential,
\be
d\equiv - \frac{\partial {\cal L_{\rm D7}}}{\partial A_\tau'} =
2\pi^2 N_f T_{D7}(2\pi\alpha')\,\frac{\rho^3}{2}\frac{\tilde
  h^2}{h}\,\cos^3\theta\,\frac{2\pi\alpha' F_{\tau\rho}+B_{\tau\rho}}
{\sqrt{1+\rho^2\theta'^2+2(2\pi\alpha'
  F_{\tau\rho}+B_{\tau\rho})^2\frac{\tilde h}{h^2}}}.
\ee 
This is a constant of motion. It is useful to define the
dimensionless combination 
\be
\tilde{d}\, \equiv\,  d\,(2\pi\alpha')^{-1}\left(2\pi^2 N_f T_{D7}\,r_H^3\right)^{-1}\,=\,
\frac{8}{\sqrt\lambda\,N_f\,N}\frac{d}{T^3}.
\label{tilded}
\ee
The solutions can be characterized by this dimensionless density
which, as we will see below, is restricted to take values less than
unity. Solving algebraically for the combination 
$(2\pi\alpha'F+B)$, using the above equations we obtain
\be
B_{\tau\rho}+2\pi\alpha'F_{\tau\rho}= \tilde d
\,\frac{2h\,\sqrt{1+\rho^2\theta'(\rho)^2}}{\sqrt{\tilde h 
\left(\tilde h^3\frac{\rho^6}{r_H^6}\cos^6\theta - 8 \tilde d^2\right)}}.
\label{bplusf}
\ee
For solutions that reach the horizon of the black hole ({\it i.e.}
which wrap the cigar $D_2$), we deduce from the above expression that 
\be
|\tilde d|\,\leq \,\cos^3\theta\big|_{\rho=r_H}.
\ee
The explicit relation between the density $\tilde d$ and the chemical potential $\mu_I$
follows from integrating Eq.(\ref{bplusf}) 
\be
\alpha-\frac{\mu_I}{T}= \sqrt\lambda \,\tilde d \,\int_{1}^\infty dy\frac
{(1-y^{-4})\sqrt{1+y^2\theta'(y)^2}}
{\sqrt{(1+y^{-4})\left((1+y^{-4})^3y^6\cos^6\theta-8\tilde d^2\right)}}.
\label{muimplicit}
\ee
Here, the dimensionless variable $y=\rho/r_H$.

The D7-brane configurations that reach the horizon are referred to as
``black hole embeddings''. Analysis of the spectral functions of
various fluctuations around such solutions reveals a continuous,
gapless spectrum \cite{mst,hoyos}
leading to the interpretation that these describe a
phase where the ``mesonic'' fluctuations have melted in the high
temperature plasma.

\subsubsection{The action}
 Evaluated on the solution for $B+2\pi\alpha' F$, the unrenormalized D7-brane
 action is
\be
S_{D7} = \lambda N_f N \frac{T^3}{64}
\int d^3x\int_{1}^\infty dy \,y^6 (1-y^{-4})(1+y^{-4})^{5/2}
\frac{\cos^6\theta\sqrt{1+y^2
 \theta'(y)^2}}{\sqrt{(1+y^{-4})^3y^6\cos^6\theta-8\tilde d^2}}.
\ee
Note that this action cannot be varied to obtain the correct equations
 of motion  for $\theta(y)$. The equation of motion for $\theta$ follows as usual by
 varying (\ref{dbiansatz}) or by using a different action where we
 trade $A_\tau$ for $F_{\tau\rho}$ as the independent variable, 
 the equation for the latter being algebraic. This is achieved by the
 Legendre transform,
\be
S_{D7}\rightarrow S_{D7}^{LT} = S_{D7} - d \int \beta \,d^3x \int
 d\rho \left(F_{\tau\rho}+\frac{B_{\tau\rho}}{2\pi\alpha'}\right).
\ee
The resulting Legendre transformed action can be evaluated on the
 solution for $F$, to yield
\be
S_{D7}^{LT}= \lambda N_f N\frac{T^3}{64}\int d^3x \int_1^\infty dy
 \, y^3 \,h\tilde h\,\sqrt{1+y^2\theta'(y)^2}\left(1-\frac{8\,\tilde
 d^2}{\tilde h^3 \,y^6\,\cos^6\theta}\right)^{1/2}\cos^3\theta\,.
\ee

It is easily verified that the corresponding expressions for real densities
and real chemical potential can be obtained by the replacement $\tilde
d \to i \tilde d$.
\subsubsection{$m_q=0$: constant solutions}
It is straightforward to see that the DBI action admits constant solutions with 
$\theta =0$ for a range of densities $\tilde d$. Since the constant
solutions extend all the way to the horizon, they describe a phase of melted
mesons. They also correspond to having 
vanishing masses for the 
fundamental hypermultiplets. In general, the large $\rho$ asymptotics of
$\theta(\rho)$ can be easily seen to be
\be
\theta(\rho)= \frac{\theta_{(0)}}{\rho}+ 
\frac{\theta_{(2)}}{\rho^3}+\ldots \,,
\label{asymp1}
\ee
where the coefficients $\theta_{(0)}$ and $\theta_{(2)}$ are related
to the quark mass and condensate respectively. 
According to the AdS/CFT dictionary, this boundary behaviour signals
the fact that $\theta$ is dual to a dimenson $3$ operator, namely, the
quark bilinear. From the earlier description of the D3-D7 system and
from Eq.(\ref{sixd}), the asymptotic value of $\rho\sin \theta$ is the
D3-D7 separation, which is the quark mass {\it i.e.},
${\theta_{(0)}}= 2\pi \alpha' \,m_q$. The coefficient of the
subleading term, namely $\theta_{(2)}$, determines the VEV of the same
operator, {\it i.e } the quark condensate
$\langle \Sigma_m \rangle=\langle\tilde \psi_i \psi_i\rangle+\ldots$. 

It follows then that the solution with $\theta=0$ has
$m_q=0,\,\langle\tilde \Sigma_m \rangle=0$. When the slipping
angle $\theta$ is vanishing, the
D7-brane wraps the equatorial $S^3$ inside the five-sphere in the
geometry. 
For this solution, 
the contribution of the flavours to the effective potential for
$\alpha$ can be computed analytically\footnote{The 
same expressions were obtained in \cite{obannon} for the zero mass
  `black hole embedding' solutions for real baryonic
  chemical potential.}. From Eq.(\ref{muimplicit}) we find 
\be
\frac{1}{\sqrt\lambda}\left(\alpha-\frac{\mu_I}{T}\right)\,\,=\,\,
\frac{\tilde d}{2}\int_1^\infty d\hat y \frac{1}{\sqrt{\hat y ^6-\tilde d^2}}\,\,=\,\,
\frac{\tilde d}{12} \,|\tilde d|^{-2/3}\, B_{\tilde
  d^2}\left(\frac{1}{3},{\tiny{\frac{1}{2}}}\right),
\label{betamu}
\ee
where $B_z(a,b)$ is the incomplete beta function and the integration variable $\hat y$ 
is related to $y$ as $\hat y^2=\frac{1}{2}(y^2+y^{-2})$. The action is
formally divergent and requires renormalizaton which is achieved
by a simple subtraction for the constant solutions,
\bea
S_{D7}^{\rm ren} &=& \lambda N_f N\frac{T^3}{16}\int d^3x
\int_1^\infty d\hat y \,\left(\frac{\hat y^6}{\sqrt{\hat y^6-d^2}} -
  \hat y^3\right)\label{betaaction}\\\nonumber
&=& \lambda N_f N \frac{T^3}{96}\int d^3 x\,\left[ |\tilde d|^{4/3}\,B_{\tilde
    d^2}\left(-\frac{2}{3},\frac{1}{2}\right)+\frac{3}{2}\right]\,.
\eea
The effective potential induced by the flavours can be found by
eliminating $\tilde d$ from equations (\ref{betamu}) and
(\ref{betaaction}). It is important to note that since $\alpha$ is a
phase defined modulo $2\pi$, 
and likewise, the imaginary chemical potential $\mu_I$ is also a periodic
variable, the combination $\alpha-\mu_I/T$ is bounded. 
\begin{figure}[h]
\begin{center}
\epsfig{file=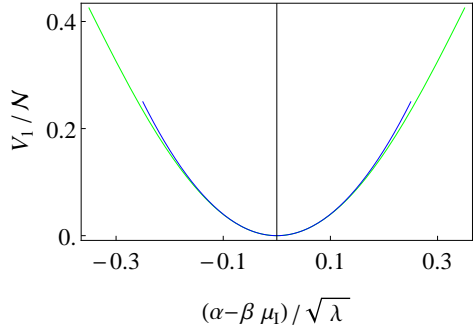, width =2.5in}
\end{center}
\caption{ \small{The potential for $\alpha$ from the flavours, as a function of
  $(\alpha-\mu_I/T)/\sqrt{\lambda}$. The curve in green is the full potential
  implied by the DBI action. The curve in blue is the
  quadratic piece that is consistent to keep in the
  $\lambda\rightarrow\infty$ limit. The normalization constant ${\cal
  N} = N_f N T^3 \lambda /16$.}} 
\end{figure}

In the
strong coupling limit $\lambda \gg 1$, we must therefore have that 
\be
\tilde d = \frac{4}{\sqrt\lambda}(\alpha-\mu_I/T)+{\cal O}(1/\lambda),
\ee
by inverting Eq.(\ref{betamu}). Since $|\tilde d| \ll 1$ at strong
coupling, we can expand the action (\ref{betaaction}) for small
$\tilde d$ to obtain the contribution to the effective potential from
the flavours,
\be
V_f= N_f N T^4 \,\left(\alpha-\frac{\mu_I}{T}\right)^2
+ {\cal O}(1/\sqrt{\lambda}).\qquad \alpha,\frac{\mu_I}{T} \in [-\pi,\pi],
\ee
which is independent of $\lambda$ at the leading order in the strong
coupling limit.
The full effective potential at strong 't Hooft coupling with
$N_f\ll N$ in the deconfined phase is then
\be
V_{\rm eff} = \,V_A+V_f\,=\,\min_{r\in {\mathbb Z}}\,4\pi^2 N^2
\frac{T^4}{\lambda}\left(\alpha-\frac{2\pi r}{N}\right)^2 +  N_f N
T^4\, \left(\alpha-\frac{\mu_I}{T}\right)^2.
\label{zeromasspot}
\ee
From this effective potential it is obvious that the degeneracy of the
${\mathbb Z}_N$ vacua has been lifted by the massless fundamental flavour
fields. Curiously, while the flavour contribution is ${\cal O}(1/N)$
suppressed compared to the leading term in the large $N$ limit, it is
unsuppressed by powers of the 't Hooft coupling $\lambda$ in the strong
coupling limit. The order of limits is, of course, unambiguous in the
present context since $N$ is taken to infinity first, keeping the 't
Hooft coupling fixed and large.
\begin{figure}[h]
\begin{center}
\epsfig{file=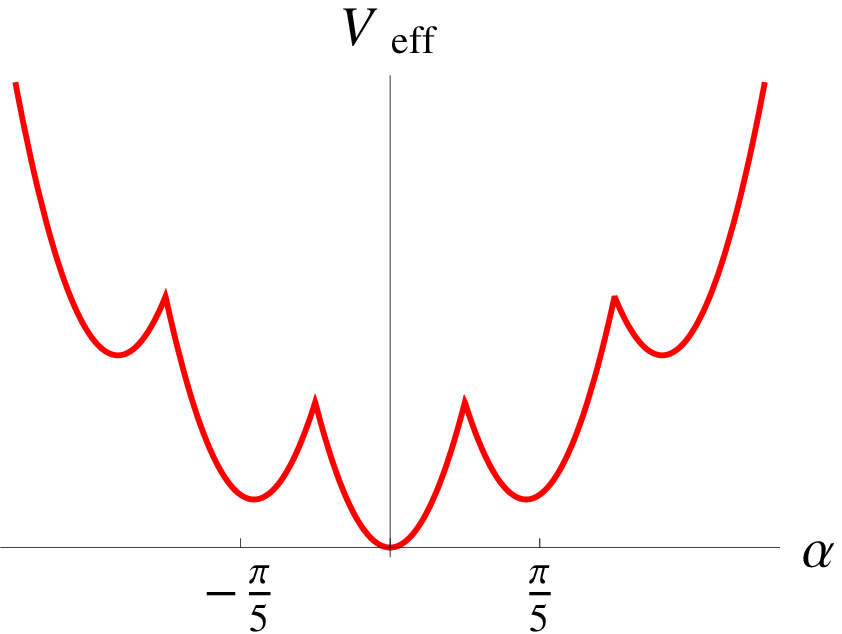, width =1.5in}
\hspace{0.2in}
\epsfig{file=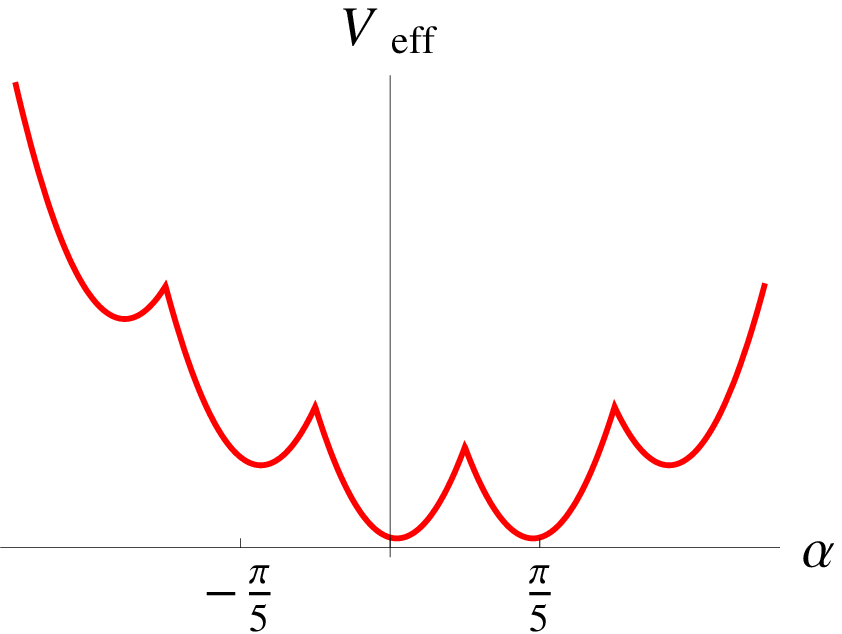, width =1.5in}
\hspace{0.2in}
\epsfig{file=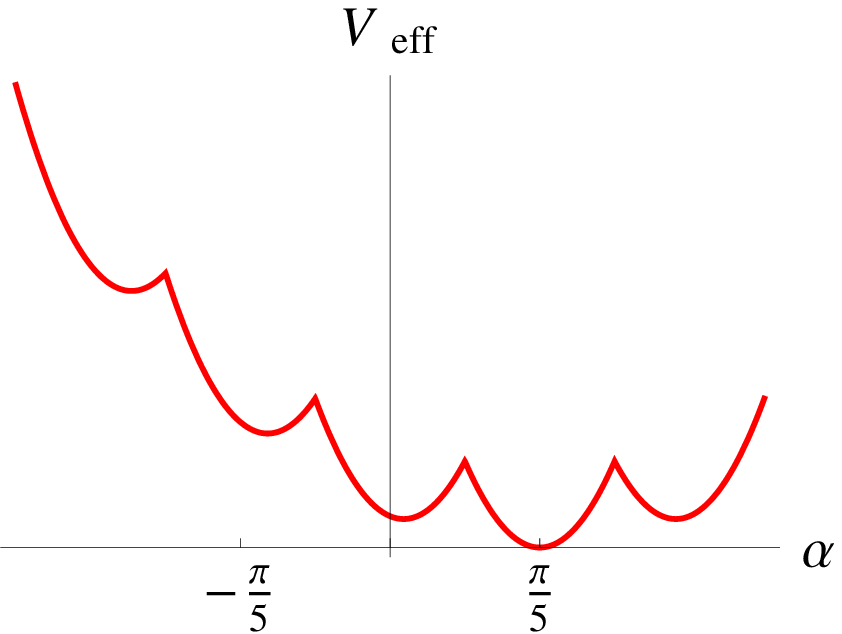, width =1.5in}
\end{center}
\caption{ \small{The full effective 
potential as a function of $\alpha$ from the flavours for three
different values of $\mu_I/T$. As $\mu_I/T$ is increased from
left to right, the local minima of the effective potential compete,
resulting in a first order phase transition.}}
\label{lifting}
\end{figure}

There are $N$ local minima, but a unique global minimum at
$\alpha=0$ when $\mu_I=0$. As ${\mu_I}$ is
increased smoothly from zero and approaches $\mu_I=\pi/N$, two
neighboring minima become degenerate and any further increase in
$\mu_I$ results in a first order phase transition from the
$\alpha=0$ vacuum to the vacuum with $\alpha=2\pi/N$ (see Fig.(\ref{lifting})).
\\
\\
{\bf \underline{Free energy:}} As the phase transitions at $\mu_I/T = (2
r-1)\frac{\pi}{N}$ for $r\in {\mathbb Z}$ are first order, it is useful to
compute the free energy as a function of the imaginary chemical
potential $\mu_I$. The free energy (the grand potential) density can be
calculated by minimizing the effective potential
Eq.(\ref{zeromasspot}) with respect to $\alpha$ for all $\mu_I$ and we
find
\bea
F[\mu_I]\Big|_{\lambda\gg 1}&=&\,N_f N \,T^4\, 
\left(1+\frac{\lambda}{4\pi^2}\frac{N_f}{N}\right)^{-1}
\,\min_{r\in {\mathbb Z}} 
\left(\frac{\mu_I}{T}-\frac{2\pi r}{N}\right)^2\,,\\\nonumber
&\simeq& N_f N T^4  \,\min_{r\in {\mathbb Z}} 
\left(\frac{\mu_I}{T}-\frac{2\pi r}{N}\right)^2\,.
\eea
While the quadratic behaviour is simply a consequence of large
$N$, it is interesting to compare the coefficient with the result at
weak coupling and zero hypermultiplet mass Eq.(\ref{fweak}). Since we
have $2 N_f$ complex scalars and $2N_f$ Weyl fermions, the free energy
at weak coupling (and $m_q=0$) is
\bea
F[\mu_I]\Big|_{\lambda\ll 1} = \frac{1}{2}\,N N_f T^4\,\min_{r\in {\mathbb Z}} 
\left(\frac{\mu_I}{T}-\frac{2\pi r}{N}\right)^2\,.
\eea
We observe that the strong and weak coupling results differ by a factor of 2.

\subsubsection{$m_q\neq 0$ non-constant solutions}
Solutions with non-zero quark mass also happen to be non-constant
and these need to be understood numerically. For smooth
solutions that get to the horizon from the boundary, we require that
the $S^3\subset S^5$ wrapped by the D7-brane remains non-vanishing
along the solution.  As
is well-known, there are two classes of non-constant D7-brane solutions: those
that get all the way to the horizon (the so-called ``black-hole
embeddings''), and those that end smoothly before they get to the
horizon. For both classes of solutions, the quark mass
and the condensate  can be read off from the asymptotic
behaviour (\ref{asymp1}) near the boundary of $AdS_5$. 
Specifically, in terms of the dimensionless variable $y=
\rho/r_H$, the boundary behaviour of the slipping angle is
\be
\theta(y) = \frac{2 }{\sqrt\lambda }\,\left(\frac{m_q}{T}\right)\,\frac{1}{y} +
\frac{\theta_{(2)}}{\, \pi^3 T^3 R^6}\,\frac{1}{y^3}+\ldots.
\ee
Note that it is only the ratio $m_q/T$ which is relevant here, since the theory
without hypermultiplets is the conformal ${\cal N}=4$ theory without
an intrinsic scale. We thus define the natural dimensionless 
ratio that characterizes these solutions,
\be
\frac{M}{T} \equiv \frac{2 m_q}{\sqrt\lambda T},
\ee
where the mass scale $M = 2m_q/\sqrt\lambda$ is the characteristic scale of 
meson bound states at zero temperature (and strong coupling) \cite{kruczenski}.

The Euclidean action for these solutions requires
careful renormalization via subtractions to yield finite results. The
correct procedure of holographic renormalization has been developed for probe
Dp-branes in AdS spacetime \cite{karchskenderis,andythesis}. The
result of this procedure in the present context is that the
renormalized action is defined as an integral over $y$ from $y=1$ (the
IR) to $y=\Lambda$ (a UV cutoff) along with counterterms defined
locally at $y=\Lambda$,
\be
S_{D7}^{\rm ren} = S_{D7}\big|_\Lambda - \lambda N_f N\frac{T^3}{64}\left(
\frac{1}{4}\Lambda^4 -\frac{1}{2}\Lambda^4
\theta(\Lambda)^2+\frac{5}{12}\Lambda^4\theta(\Lambda)^4\right).
\ee
The first two counterterms are divergent while the third is a finite
subtraction. Further counterterms are necessary if the boundary is not
flat.
\\\\
{\bf \underline{Black-Hole embeddings:}} 
Solutions extending all the way to the
black hole horizon, often referred to as ``black-hole embeddings'' of
the D7-brane, satisfy $\theta'(y=1) =0$ and $0\leq \theta(y=1) <
\frac{\pi}{2}$. Thus for these embeddings, the $S^3$ remains
blown-up all along the solution, while the $S^1$ (the thermal circle)
shrinks at the horizon, so the D7-brane caps off
smoothly. By varying $\theta(y=1)$ we can explore different values of
the hypermultiplet mass $M/T$. These Euclidean solutions can exist
only if 
\be
|\tilde d| \leq \cos^3\theta\big|_{y=1}.
\label{condition}
\ee
The black hole embeddings describe a high temperature phase where
mesons have melted.

For a fixed density $\tilde d$, solutions with different
hypermultiplet masses can be found by dialling $\theta(y=1)$ provided
the condition (\ref{condition}) is satisfied. As in the massless case
analyzed above, we will obtain (numerically) the effective potential
as a function of $(\alpha-\mu_I/T)/\sqrt\lambda$. We summarize the
properties of the solutions below.

\begin{figure}[h]
\begin{center}
\epsfig{file=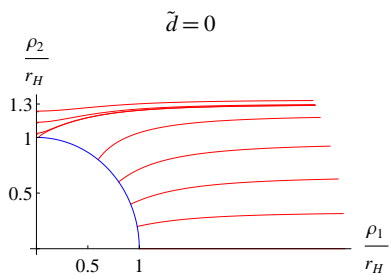, width =2in}
\end{center}
\caption{ \small{The well known transition at zero density, from black
  hole embeddings (falling into the black hole) to Minkowski
  embedding probe D7-branes placed away from the horizon.}}
\label{critical} 
\end{figure}

\begin{itemize}
\item {The solutions for finite density $0 < |\tilde d| \leq 1$, are
all of the type described above and are 
qualitatively similar to black-hole embeddings
found earlier for vanishing baryonic chemical potential
\cite{babington,mateos1}.  They are qualitatively
distinct from the configurations with real chemical potential
\cite{mateos2,ghoroku2,obannon,mateos4}. We elaborate on this below.}

\item{For any given density $|\tilde d|$ (between 0 and 1) 
there is a fixed value of $M/T$, above which solutions cease to
exist. The allowed range of masses decreases with increasing $|\tilde
d|$ (Figure \ref{range}). 
 \begin{figure}[h]
\begin{center}
\epsfig{file=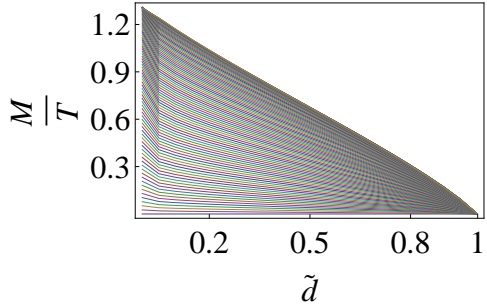, width =2.5in}
\end{center}
\caption{ \small{ The values of $\frac{M}{T}$ and the density
    variable $\tilde d$ (conjugate to imaginary $\mu$) 
for which the black hole embeddings or melted meson solutions exist.
}}
\label{range} 
\end{figure}
When the density $\tilde d$ approaches unity, all allowed
solutions bunch up near the zero mass constant solution. 
This is illustrated in Figure \ref{blackhole} where we plot 
solutions in the $\rho_1-\rho_2$ plane ($\rho_2 =\rho\sin\theta $ and
$\rho_1=\rho\cos\theta$). In contrast to this picture, for a real
chemical potential, it was observed (e.g. in \cite{mateos2,obannon}) 
that black hole embeddings can exist at all possible values of the ratio
$M/T$. 
For large values of the mass (or low
temperatures), these solutions could be viewed as a probe D7-brane
with strings (quarks) attached which drop into the black hole
horizon. The absence of these kinds of solutions substantially alters
the phase diagram for imaginary chemical potential.}  

\begin{figure}[h]
\begin{center}
\epsfig{file=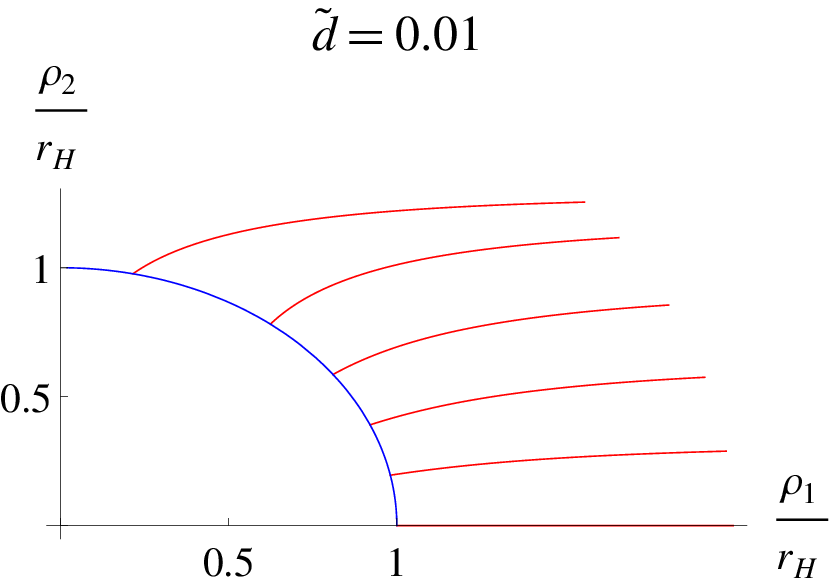, width =1.7in}
\hspace{0.1 in}
\epsfig{file=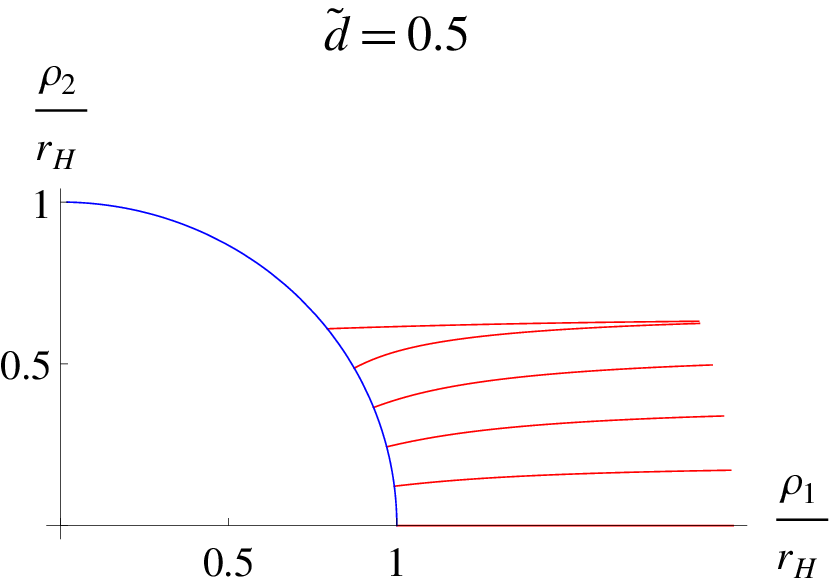, width =1.7in}
\hspace{0.1 in}
\epsfig{file=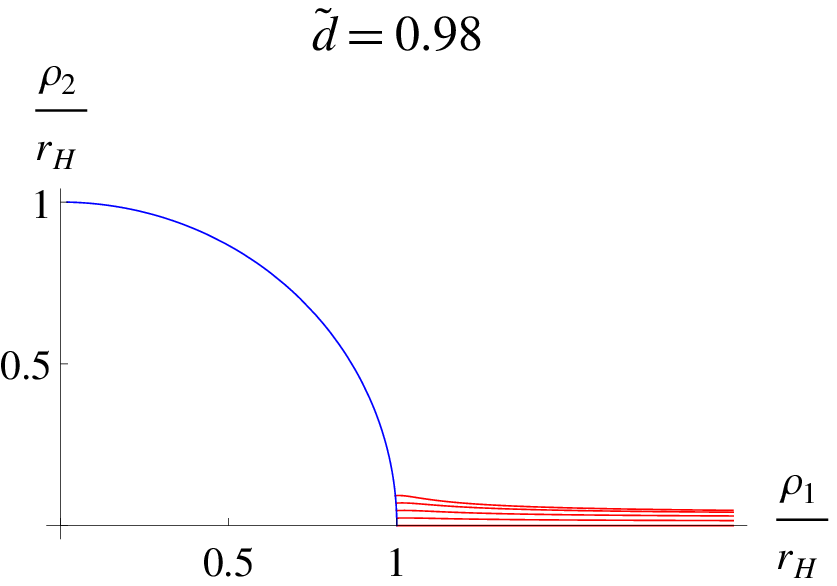, width =1.7in}
\end{center}
\caption{ \small{D7-brane solutions extending to the horizon for
    $\tilde d=0.01$, $0.5$ and $0.98$. For each
    density we have plotted a family of solutions with increasing masses
    up to the maximum value allowed for that density. The
    asymptotic distance from the $\rho_1$ axis gives the ratio
    $M/T$. The black hole horizon is 
the blue curve $\rho_1^2+\rho_2^2=r_H^2$.
}}
\label{blackhole} 
\end{figure}

\item{The largest value of the ratio $\frac{M}{T} = \frac{2 m_q}{\sqrt\lambda T}$ 
for which the black hole embedding exists is 
$\approx 1.30$ and this occurs when $\tilde d=0$ and is the
critical embedding solution (Figure \ref{critical}).}

\begin{figure}[h]
\begin{center}
\epsfig{file=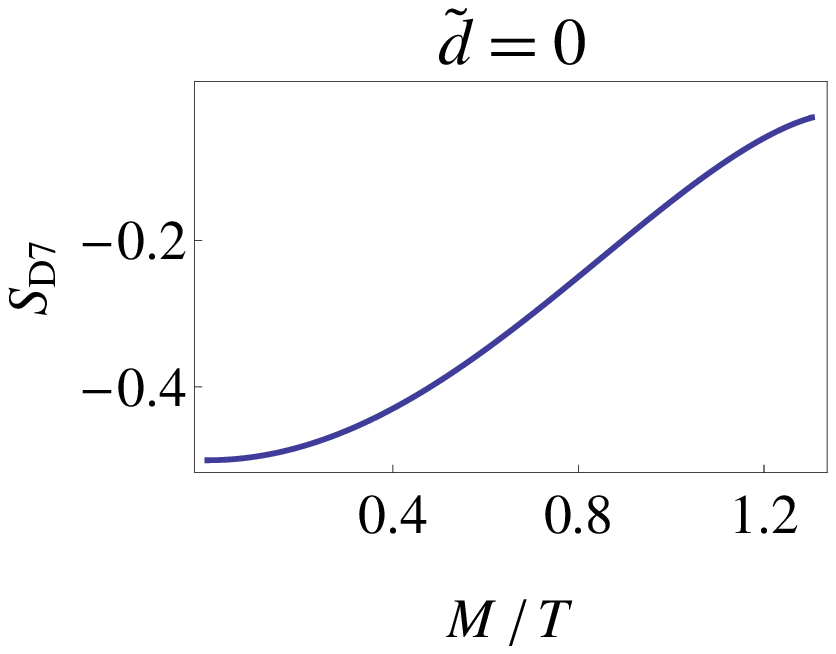, width =1.5in}
\hspace{0.2 in}
\epsfig{file=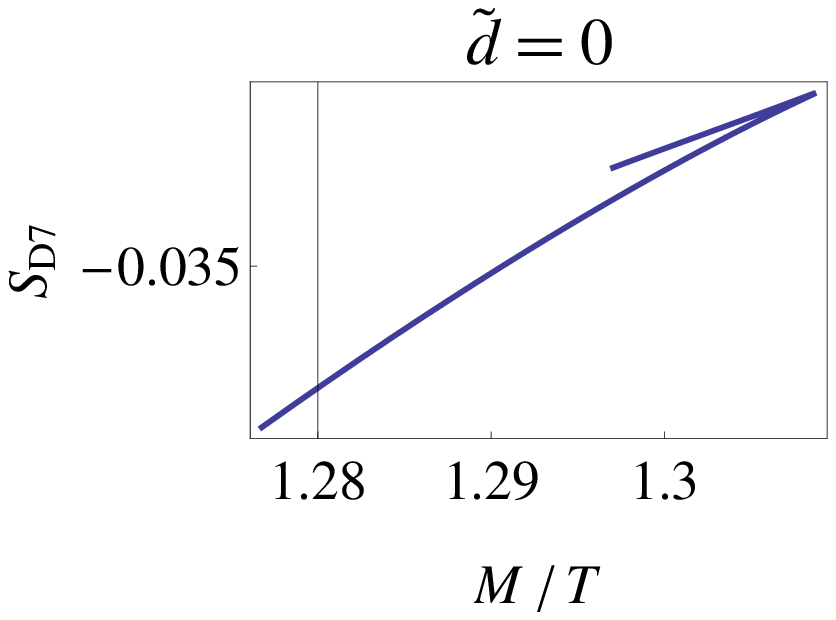, width=1.5in}
\hspace{0.2 in}
\epsfig{file=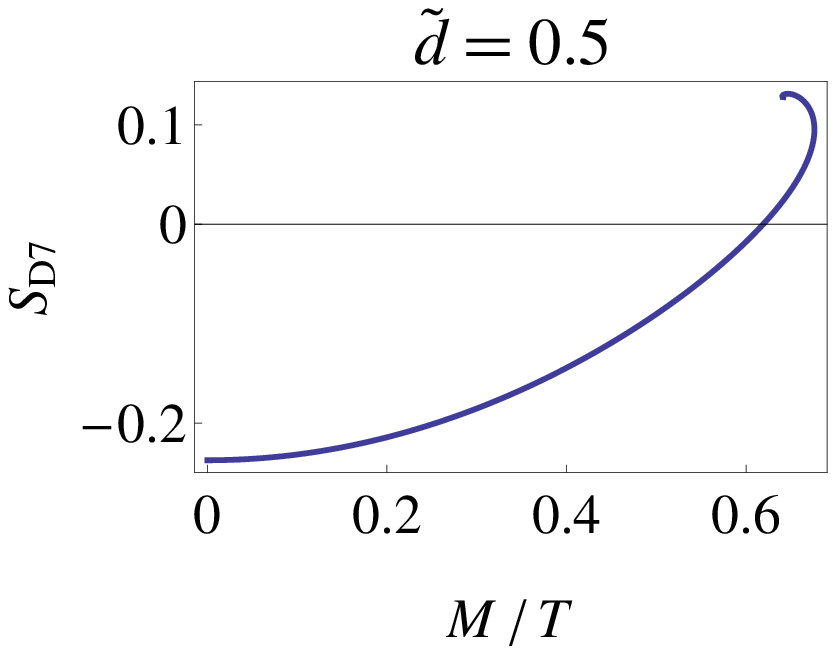, width =1.5in}
\end{center}
\caption{ \small{The action as a function of $M/T$ for $\tilde d =0$ and
    $\tilde d=0.5$.}}
\label{action0} 
\end{figure}

\item{By evaluating the action and the expectation value of the quark condensate operator
    $\langle \Sigma_m\rangle$, for a given density $\tilde d$, 
 we may see that for large enough masses, there exist
    two classical  
solutions with the same mass but different values for the action and 
condensate (Figure \ref{action0}). This feature, well-known at zero
    density, also appears for non-zero densities conjugate to an 
    imaginary chemical potential. The solution with
    the lower action is then chosen to be the physical one.}
\end{itemize}

\begin{figure}[h]
\begin{center}
\epsfig{file=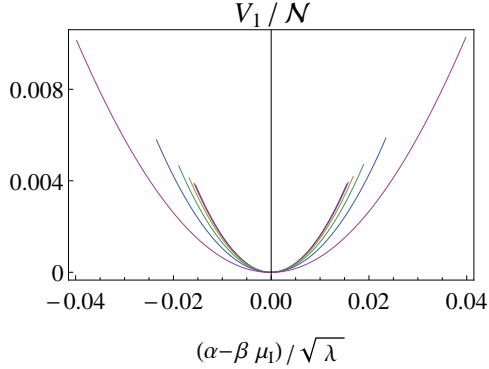, width =2.5in}
\end{center}
\caption{ \small{The effective potential from flavours for increasing
    values of $m_q$, for $M/T =0$, 
$0.24$, $0.48$, $0.72$, $0.96$ and $1.2$. The potential tends to flatten out 
as $M/T$ approaches the critical value of $1.3$.
}}
\label{massivepot}
\end{figure}

\begin{figure}[h]
\begin{center}
\epsfig{file=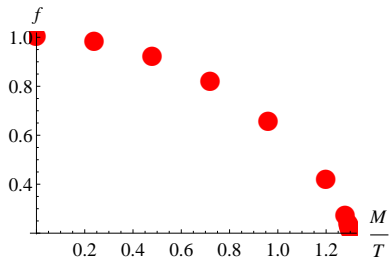, width =2in}
\end{center}
\caption{ \small{The dimensionless coefficient $f$ in the quadratic 
    effective potential $V_f$ due to hypermultipltes.
}}
\label{fm} 
\end{figure}

{\bf \underline{The effective potential:}}  The contribution to the
effective potential for the phase of the Polyakov-Maldacena loop $\alpha$, can
be found numerically (Figure \ref{massivepot}). 
However, we can already be fairly precise about
the form of this contribution due to Eq. (\ref{muimplicit}): since
$\alpha$ and $\mu_I/T$ are bounded, we must have that $|\tilde d|
\sim {\cal O}(1/\sqrt\lambda) \ll 1$. Notice that this scaling of
$\tilde d$, actually implies that the dimensionful density $d\sim N_f
N T^3$, which is a natural scaling for $N_f$ quark flavors. In taking
$\tilde d \sim {\cal O}(1/\sqrt\lambda)$, we are also assuming that the integral
in Eq. (\ref{muimplicit}) is well-behaved as $\tilde d \to 0$
and we can check numerically that this is the case for all hypermultiplet
masses in the range $0 \leq \frac{M}{T}\lesssim 1.30 $ corresponding
to the black hole embedding 
solutions. In addition since the D7-brane action is a function of
${\tilde d}^2$, for small $\tilde d$, it can be expanded in powers of
${\tilde d}^2$. In the large $\lambda$ limit, the quadratic piece 
alone dominates and therefore we
find that the effective potential is of the form
\be
V_{\rm eff} = \,V_A+V_f\,=\,\min_{r\in {\mathbb Z}}\,4\pi^2 N^2
\frac{T^4}{\lambda}\left(\alpha-\frac{2\pi r}{N}\right)^2 +  N_f N
T^4 \,f\left(\frac{M}{T}\right)\left(\alpha-\frac{\mu_I}{T}\right)^2.
\label{genericmass}
\ee
Here $M$ is the mass scale 
$\frac{2m_q}{\sqrt\lambda}$ and $f$ is a dimensionless
function of $\frac{M}{T}$. 
The coefficient $f\left(\frac{M}{T}\right)$ is unity when $m_q=0$ and decreases
monotonically with increasing mass, eventually approaching $\approx 0.2$  
as $\frac{M}{T} $ is dialled to its maximum value of approximately 
$1.3$ (Figure \ref{fm}).
We should emphasize that in deriving the effective potential above, in
the large $\lambda$ and large $N$ limit, we only needed to use
solutions with small $\tilde d$, strictly only those in the vicinity of 
$\tilde d\sim{\cal O}(1/\sqrt\lambda)$. When we turn to the low
temperature regime below, we will see that this is actually crucial
for ensuring a consistent picture of the transition between high and
low temperatures.

To summarize, for all values of the parameter $\frac{M}{T}$ for which
black hole  
embeddings of the D7-brane exist, with an imaginary chemical
potential, first order Roberge-Weiss transitions will occur at $\frac{\mu_I}{T}= (2r-1)\frac{\pi}{N}$, $r\in {\mathbb Z}$. This means
that, for fixed hypermultiplet mass $m_q$, the RW transition
will occur for temperatures $T \gtrsim 2 m_q/(1.3\sqrt\lambda)$. We
will determine the value of this temperature, 
representing the RW endpoint, more precisely below.

The monotonic decrease in $f(M/T)$ with decreasing temperature
(or increasing mass $m_q$) is in accord with intuition from weak
coupling. The quadratic potential for $\alpha$ can be interpreted as a
thermal contribution to the mass of the mode $\alpha$ from the
flavours. That this thermal correction should decrease as the flavours
are made heavier (eventually decoupling for infinite mass),  
appears intuitively to be consistent. At weak coupling, it
would be natural to identify the mass of the mode $\alpha$ with the
Debye mass (inverse electric screening length). It is unclear whether
any such interpretation should be possible at strong coupling. Finally, we can
again write the free energy as a function of $\mu_I$, (taking
$\lambda N_f/N \ll 1$),
\be
F[\mu_I] =  N_f N T^4 \, f\left(\frac{m_q}{2\sqrt\lambda T}\right)\,
\min_{r\in {\mathbb Z}}\,(\alpha-\mu_I/T)^2.
\ee
Given our numerical results for $f$, we can say that near $m_q=0$, 
$f \approx 1 - f''(0)\frac{m_q^2}{8\lambda T^2}$. In contrast, in the
weakly coupled theory the coefficient in Eq. (\ref{fweak}) 
(with $n_{f}=\tilde n_{f} =2N_f$) decreases
{\em linearly} with mass, for small enough mass, and is independent of
the weak 't Hooft coupling at leading order
\be
\frac{1}{6}\left(\,f_{\rm bose}(m_q/T)+ f_{\rm
    fermi}(m_q/T)\,\right)\approx \frac{1}{2}- 
\frac{|m_q|}{T}\, \frac{\pi}{18}+\ldots
\ee
The origin of the linear term can be traced to the pole in the
Bose-Einstein distribution for free massless bosons at zero momentum. 
The absence of
a linear term at strong coupling is perhaps indicative of the strong
dressing undergone by the perturbative scalar degrees of freedom. It
is also worth noting that at strong coupling, the natural dimensionful 
parameter is the meson mass scale $M$, rather than the hypermultiplet
quark mass $m_q$. 

{\bf \underline{Low temperatures/Large masses, $ M \gtrsim 1.3$:}}
The critical black hole embedding of the D7-brane has the property
that the slipping angle $\theta$ of the solution approaches
$\frac{\pi}{2}$ at the horizon and the corresponding $S^3$ wrapped by
the probe brane shrinks. 
Hence, the only solutions with masses
higher than $\frac{M}{T} \approx 1.3$ are those where the $S^3$ wrapped
by the D7-brane shrinks before the D7-brane gets to the horizon. For
these embeddings the thermal $S^1$ remains non-vanishing while the
$S^3$ shrinks smoothly before the brane gets to the horizon. Since the
$S^3$ shrinks at the tip of such a solution, we are forced to have
$\tilde d=0$ due to Eq.(\ref{condition}) and hence
\be
{F_{\tau\rho}+2\pi\alpha'B_{\tau\rho}}=0.
\ee
The D7-brane action has no dependence on either $\alpha$ or
$\mu_I$. 
Hence there is no potential induced by flavours for the phase
$\alpha$ of the Polyakov-Maldacena loop: $V_f(\alpha-\mu_I/T)$
vanishes identically  
and the physics is completely smooth as a function $\mu_I$. Note that
despite the fact that the field strength $F$ is fixed by $B$, we can always
add a constant to the world-volume gauge potential $A_\tau$, which will
not change the field strength, and can be interpreted as a chemical
potential $\mu_I$ in the boundary field theory. Hence, in the grand
canonical ensemble, at low temperatures ($T\ll M$), for every
$\mu_I$ the only solution possible is the $\tilde d =0$ Minkowski
embedding representing unmelted mesons. The situation is quite
distinct from the case of real chemical potential \cite{mateos4}
wherein, at fixed low temperatures, there is a transition from
Minkowski embedding solutions to the spiky black hole embeddings as
the chemical potential is increased.

\begin{figure}[h]
\begin{center}
\epsfig{file=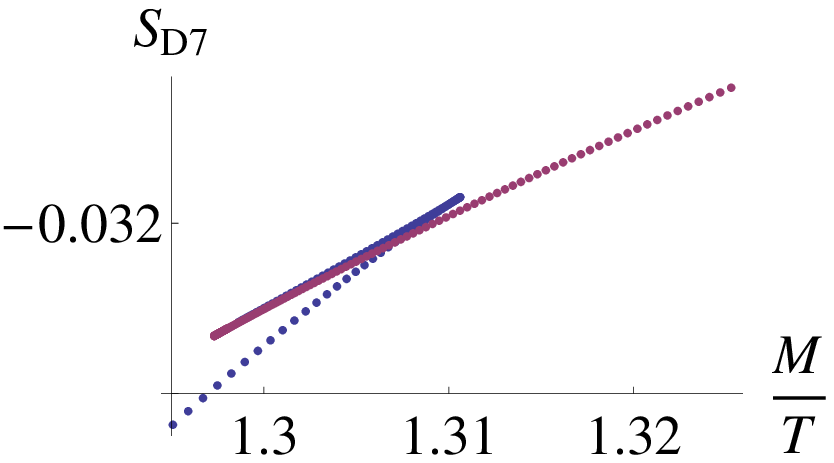, width =2.2in}
\hspace{0.5in}
\epsfig{file=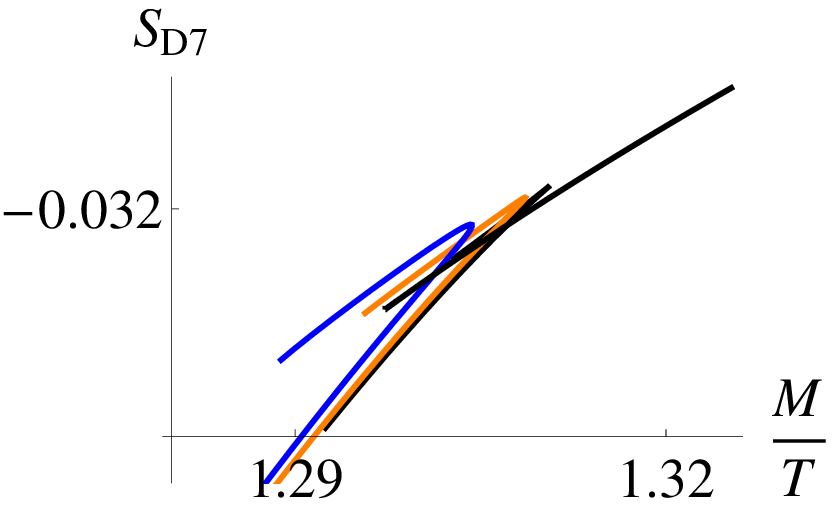,width=2.2in}
\end{center}
\caption{ \small{
\underline{Left}: The action 
for $\tilde d=0$, D7-brane embeddings. As the ratio
$M/T$ is increased, just before the black hole embedding (melted
meson) ceases to exist, there is a   transition to the family of so-called 
``Minkowski embeddings'' corresponding to unmelted mesons. 
\underline{Right:} In the second figure we see that the
action (grand potential) 
for black hole embeddings with $\tilde d > 0$, (0.004 (orange)
and 0.008 (blue)), is always larger than the action for the $\tilde
d=0$ solutions (black). There is no Minkowski
embedding solution with $\tilde d \neq 0$. 
}}
\label{transition} 
\end{figure}

The
solutions with $\tilde d=0$ are the same as those originally obtained in
\cite{babington, mateos1, karch1}. The
transition from the black hole embeddings to the second class of
solutions involves a topology change: the first category of solutions
exhibits a shrinking $S^1$ and a finite $S^3$, while the second 
has the $S^3$ shrinking in the interior. The transition between these
is well-known and is a first order phase transition as may be inferred
from plotting the action as a function of $M/T$ (Figure
\ref{transition}) for the solution with $\tilde d =0$.
The phase transition between the two classes of solutions is a
``meson-melting'' transition. At low enough temperatures (or large
enough hypermultiplet mass), the only allowed D7-brane embeddings are
the unmelted mesons with zero density $\tilde d=0$. The spectrum of
fluctuations about this solution exhibits a mass gap and a discrete
spectrum \cite{kruczenski} at zero temperature. The mass gap in the
meson spectrum is  
\be
M_{\rm gap} = 4\pi\frac{m_q}{\sqrt\lambda} = 2\pi M.
\ee

We also see that as the density $\tilde d$ 
is increased from zero, for small enough densities $\tilde d \ll 1$, 
the black hole embeddings with $\tilde d \neq 0$
coexist with the zero density $\tilde d=0$ Minkowski embedding
solution for a range of temperatures (Figure \ref{transition}). This
will be important for determining the phase diagram and the shape of the
phase boundaries in the grand canonical ensemble. With increasing
density, there is a critical value of $\tilde d$, ($\gtrsim 10^{-2}$) 
beyond which the curves for black hole embeddings (as in Figure
\ref{transition}) cease to intersect the Minkowski solution at $\tilde
d=0$. They turn back and terminate at progressively smaller values of
$M/T$ (Figure \ref{action0}). The presence of such configurations in
the grand canonical ensemble could make the low temperature region of
the phase diagram inaccessible. However, it should be clear from the
preceding discussions that these configurations are essentially
removed as a consequence of the Roberge-Weiss transition. The RW
transitions occur at values of $(\alpha-\mu_I/T)$ that are
parametrically suppressed by $1/N$ corresponding to parametrically
small values of $\tilde d$. The configurations with such low densities
will always be of the kind depicted in Figure \ref{transition}.

\section{Phase diagram}
\label{sec:phase}

The above analysis of the dominant configurations at different densities 
and temperatures now allows us to determine the phase diagram of the 
theory as a function of the temperature/meson mass scale ratio $T/M$ and 
the imaginary chemical potential $\mu_I$. We have already established that 
at low temperature the physics is independent of $\mu_I$ and $\alpha$, 
the phase of the Polyakov loop, and is dominated by the unmelted meson 
(Minkowski embeddings) configurations with vanishing density $\tilde d=0$. 
In addition, we have seen that at high temperature the dominant 
configurations are the black hole embeddings representing melted mesons 
with $\tilde d \neq 0$.  For these configurations the system experiences a 
first order phase transition at $\mu_I/T= (2r-1)\pi/N$ ($r\in {\mathbb 
Z}$), characterized by a discrete jump in the phase of the Polyakov loop. 
Specifially, $\alpha$ jumps from $2\pi(r-1)/N$ to $2\pi r/N$.

What remains is to understand the phase boundary between
the melted and unmelted  meson phases as a function of $\mu_I$.
We first recall that in the melted phase $\tilde d$ is proportional to
$(\mu_I/T-\alpha)/\sqrt\lambda$, for small enough $\tilde
d$. As depicted in Figure \ref{transition}, in the grand canonical
ensemble, the melted meson configuration with small $\tilde d \neq 0$
will then have to compete with the $\tilde d=0$ Minkowski 
embedding as the temperature is decreased. This results in a first order
transition between the two phases. It is clear from Figure \ref{transition} 
that as the density increases (for positive $\tilde d$, say), 
the transition temperature increases as well (equivalently $M/T$
decreases). 
\begin{figure}[h]
\begin{center}
\epsfig{file=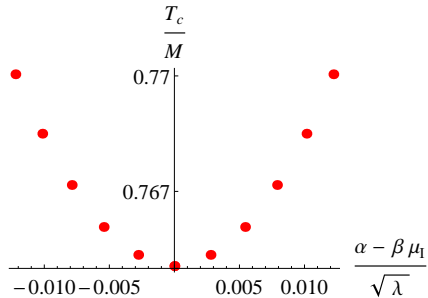, width =2.2in}
\end{center}
\caption{ \small{ The melting transition temperature at different
    values of $\mu_I$ and fixed $\alpha$.
}}
\label{phase1} 
\end{figure}
 The dependence of this transition temperature on the chemical potential 
determines the shape of the phase boundary in the $\mu_I-T$ plane. We can 
numerically obtain the melting transition temperatures $T_c(\mu_I)$ for 
different values of $\mu_I$ (corresponding to different densities $\tilde 
d$ in the melted phase). The result is shown in Figure \ref{phase1}.
 For 
\begin{equation}
\alpha= \frac{2\pi r}{N}\,,\qquad (2r-1)\frac{\pi}{N}\leq\,
\frac{\mu_I}{T} \,< \,(2r+1)\frac{\pi}{N}\,, \qquad r\in {\mathbb Z},
\end{equation}
we find that 
\be
\label{eq:Tc}
T_c(\mu_I) -T_0 = K\,\frac{M}{\lambda}\left (\frac{\mu_I} {T_c(\mu_I)} - \alpha
\right)^2+\ldots, \qquad K\approx 33.5 \\\nonumber\\\nonumber
\ee
where $T_0=T_c(0)\approx 0.77M$.
We have only kept terms to quadratic order on the right hand
side for two reasons: First, since the difference $\mu_I/T-\alpha$
is bounded, higher order terms are suppressed at strong
coupling ($\lambda \gg 1$). 
Secondly, we also have that $|\mu_I/T-\alpha|\leq
\pi/N$, so higher order terms are suppressed by powers of $1/N^2$.
At leading order in $1/\lambda$, $T_c(\mu_I)$ may therefore be replaced  by $T_0$
on the right-hand side of (\ref{eq:Tc}).

The first-order phase boundary between the melted phase at high temperature
and the unmelted phase at low temperature has the curvature as 
shown in Figure \ref{phase2}. Combining this result with the Roberge-Weiss 
lines at $\mu_I=\muRW$ yields an infinite set of points at which 
three first order transition lines meet. At each of these 
{\em triple points}, three phases labelled by distinct values of the
Polyakov loop coexist. The location of the triple points in the
$\mu_I-T$ plane is given by ($\muRW, \TRW$), where
\be
\frac{\muRW}{T_0} = (2r-1)\frac{\pi}{N}, \quad\quad\quad  \frac{\TRW}{T_0} = 1+\frac{\kappa}{\lambda}\frac{\pi^2}{N^2},
\ee
with $r\in {\mathbb Z}$ and $\kappa=KM/T\approx 43.5$.
The $N$-scaling of the
formulae may suggest that these should be  subleading
effects in the large-$N$ limit and therefore not consistently
incorporated in a classical supergravity approximation. 
However the factors of $1/N$ arise from purely
kinematical considerations, namely the symmetry of the theory under
shifts $\mu_I\to \mu_I+2\pi/N$. When $N$ is large, the RW transitions are
closely spaced, resulting in the $N$-dependence in the Roberge-Weiss
temperature. The curvature of the phase boundary in Figure
\ref{phase2} is independent of $N$, see Eq.\ (\ref{eq:curv}).

\subsection{Real chemical potential and analytical continuation}
For real values of the quark density a rich family of solutions to the
DBI equations of motion exists, after making the replacement $F\to i F$ 
(and setting $B=0$) in Eq.\ (\ref{dbiansatz}). In particular, black hole
embeddings, representing melted mesons, exist for all values of the
ratio $M/T$ in the canonical ensemble with fixed quark number density.
These include low temperature ($M\gg T$) solutions which can
be viewed as Minkowski embeddings attached to a spike consisting of
strings (quarks) falling into the black hole horizon. 

The phase diagram for real chemical potential in the grand canonical ensemble \cite{ghoroku2,
  Nakamura:2007nx, obannon, mateos4} bears little relation to Figure
  \ref{phase2}, either at low or high temperatures. With a real quark chemical
potential, there is a single line of phase transitions separating
a zero density phase from the dissociated or
melted meson phase. For small chemical potential the transition line
is first order, while at larger chemical potential the transition is expected to be a
  continuous one \cite{obannon, liu, benincasa}, terminating at $\mu=m_q$. In
fact it has been argued in \cite{liu} that for low temperatures the
transition line is actually third order and connects up with the first
order line at a tricritical point.

In the context of QCD, the main motivation behind the exploration of the phase
diagram as a function of imaginary chemical potential has been to determine the phase
structure for real chemical potential, via the use of analytic continuation from imaginary $\mu=i\mu_I$
(with $\mu^2<0$) to real $\mu$ (with $\mu^2>0$) 
\cite{deForcrand:2002ci,deForcrand:2003hx,deForcrand:2006pv, 
deForcrand:2010he,D'Elia:2002gd,D'Elia:2004at,D'Elia:2007ke, 
D'Elia:2009tm,D'Elia:2009qz,Cea:2010md}.
It is therefore 
natural to ask what aspects of the  phase diagram 
can be captured by analytic continuation in the holographic model we studied here.
This is particularly interesting given the qualitative difference in the nature of black hole
embeddings and the D7-brane solutions at fixed density.
At  large real or imaginary chemical potential the phase structures are manifestly different  due to the presence of the Roberge-Weiss transitions. However it is expected that the curvature of the meson
melting line for imaginary chemical potential, with $\mu^2<0$, is directly
related to the curvature of the first order transition line for real chemical potential, with $\mu^2>0$. 
\begin{figure}[h]
\begin{center}
\epsfig{file=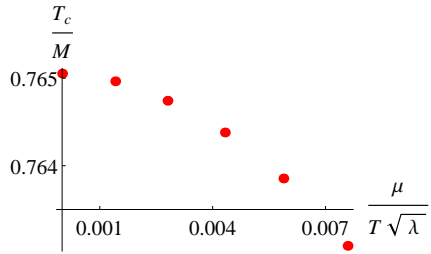, width =2.2in}
\end{center}
\caption{ \small{ The line of first order meson-melting 
transitions for real chemical potential, $\mu\gtrsim 0$.
}}
\label{realchem} 
\end{figure}
Repeating the analysis for real chemical potential we find that the
curvature of the first order line for $\mu^2>0$ is given by
(see Figure \ref{realchem}) 
\be
T_c(\mu)- T_0 \approx - 33.5 \, \frac{M}{\lambda} \frac{\mu^2}{T^2_0},
\ee
matching up with our result for imaginary $\mu$.
It is reassuring that despite the differences between individual
D7-brane probe solutions for $\mu^2\gtrless0$, the theory is 
analytic in $\mu^2$ near $\mu^2=0$.

\section{Discussion}
 In the presence of an imaginary quark chemical potential $\mu_I$ the 
D3-D7 holographic setup at strong coupling displays a first-order meson 
melting transition at $T=T_c(\mu_I)$ and a set of first-order 
Roberge-Weiss transitions at fixed $\mu_I=\muRW$ in the high-temperature 
phase $T>T_c(\mu_I)$. The first-order lines join at a series of triple 
points in the $\mu_I-T$ plane. We have determined the location of the 
triple points and the curvature of the phase boundaries, and confirmed 
that the theory is analytic in $\mu^2$ near $\mu^2=0$.

Since the adjoint degrees of freedom are always deconfined in this model, 
the quark (or meson) mass and the temperature cannot be varied 
independently in the absence of another scale. The model of Sakai and 
Sugimoto \cite{ss} would be more ``QCD-like'' in this context and it would 
therefore be interesting to compute the order of the thermal phase 
transition and the Roberge-Weiss endpoint in that model.
Recently, it has been demonstrated in QCD that the Roberge-Weiss endpoints
are triple points for two \cite{D'Elia:2009qz} or three
\cite{deForcrand:2010he} degenerate light or heavy flavours, and second
order endpoints in the 3d Ising universality class for intermediate quark
masses. It would be extremely interesting to find a 
holographic model where the triple points turn into second order 
endpoints, as the quark mass $m_q$ is varied. This could be realized
in models incorporating backreaction of flavours or in defect models
in general Dp-Dq setups \cite{benincasa}.
\\

{\bf Acknowledgements:} We would like to acknowledge an STFC rolling
grant for support. We thank Paolo Benincasa for comments on a draft
version of this paper; Owe Philipsen and Massimo d'Elia for discussions. 
\\ 

\newpage
\startappendix
\Appendix{One-loop free energy at weak coupling}
To calculate the perturbative contributions to the 
free energy at finite temperature, in the presence
of an imaginary chemical potential, we need to first understand the
perturbative effective potential for the Polyakov loop degrees of
freedom. The computation is fairly standard, so we will only
present a brief summary (we refer the reader to 
\cite{Aharony:2003sx, weiss, sbh} for
details of the theory on flat space and on $S^3\times {\mathbb
  R}$). At finite temperature, for the $SU(N)$ gauge theory, we
introduce the Wilson line around the thermal circle as
\be
U= \exp i\left(\int_0^\beta A_\tau\,d\tau\right).
\ee
Picking a gauge where the zero mode of $A_\tau$ is diagonal
\be
{\cal A}\equiv\int_0^\beta A_\tau \,d\tau \,= \,
{\rm Diag}(\alpha_1,\alpha_2,\ldots \alpha_N),\,\qquad
\sum_{i=1}^N \alpha_i = 0 \, {\rm mod} \, 2\pi,
\ee
we want to compute the perturbative contributions to the thermal
partition function in the presence of these VEVs. Let us suppose that
the $SU(N)$ gauge fields are coupled to $\tilde n_A$ adjoint Weyl fermions and
$n_{A}$ real scalars also in the adjoint representation of the gauge
group. In addition we allow for $\tilde n_{f}$ Weyl fermions
and $n_{f}$  complex scalars all in the fundamental representation
of $SU(N)$. The one-loop result for the effective action then has the
general form
\bea
S_{\rm 1-loop} = \hm&&
\frac{1}{2}\log {\det}\left(- {\cal D}_0^2 - \nabla^{2(v)}\right)
+ \frac{n_{A} }{2}\log\det\left(-{\cal D}_0^2 - \nabla^{2}\right)
\nonumber \\ &&
+n_{f}\,\log\det\left(-{\tilde D}_0^2 - \nabla^{2}+m^2\right)
-\tilde n_{A}\log\det\left(i\gamma^0 {\cal D}_0 + i \gamma^i
  \partial_i\right)
\nonumber \\ &&
- \tilde n_{f}\log\det\left(i\gamma^0 \tilde D_0 + i \gamma^i
  \partial_i -m_q\right). 
\eea
Here, we have allowed for a common mass $m_q$ for the fields in the
fundamental $(f)$ representation in view of the fact that the theory that we 
investigate at strong coupling also falls in this class. The background field
gauge covariant derivatives are defined as
\be
{\cal D}_0 \equiv \partial_\tau + i T \,[{\cal A},\,]\,\,,\qquad 
{\tilde D}_0 \equiv \partial_\tau + i T {\cal A}\,\,,
\ee
when acting on adjoint and fundamental fields respectively. This
one-loop effective action is then given by
\bea
&&S_{\rm 1-loop}/{\rm Vol}\label{effective}\\\nonumber
&&=\int_0^\infty \frac{dk}{2\pi^2}\, k^2\,
  \left[ \sum_{ij=1}^N\left(\,(2+ n_{A})\,{\rm Re}\,
\ln\left(1-e^{-i  \alpha_{ij}}\, e^{-\beta k }\right)
- 2 \tilde n_A\,{\rm Re}\,
\ln\left(1+e^{-i  \alpha_{ij}}\, e^{-\beta k
  }\right)\right)\right.\\\nonumber
&&\left.
+ \sum_{i=1}^N\,\left( 2 n_{f}\,{\rm Re}\,
\ln\left(1-e^{-i  \alpha_{i}}\, e^{-\beta \omega_k }\right)
-2 \tilde n_{f} \,{\rm Re}\,
\ln\left(1+e^{-i  \alpha_{i}}\, e^{-\beta \omega_k }\right)\right)
\right].
\eea
Here
\be
\alpha_{ij}\equiv \alpha_i-\alpha_j\,\,,\qquad \omega_k\equiv
\sqrt{k^2+m_q^2}.
\ee

To evaluate the free energy as a function of the imaginary quark
number chemical potential, we will consider the massless and massive
cases separately, as the former can be studied exactly at the one-loop
order. 

{\bf \underline{Massless fundamental flavours:}} 
With $m=0$ the
integrals in Eq. (\ref{effective}) can be done to yield the effective
potential for the Polyakov loop variables (assuming $0\leq \alpha_i<
2\pi$ and $\sum \alpha_i=0 \,{\rm mod}\, 2\pi$),
\bea
V_{\rm eff}\label{effpot}
=\hm&& \frac{\pi^2}{48}T^4\left[
\sum_{ij=1}^N\left\{
 (2+n_{A})\left(1-\left(\frac{\alpha_{ij}}{\pi}\Big|_{\rm mod\, 
2}-1\right)^2\right)^2 \right.\right.
\\ \nonumber &&
\left.\left.
- 2 \tilde n_A\left(1-\left(\left(\frac{\alpha_{ij}}{\pi}+1\right)
\Big|_{\rm mod \, 2} -1\right)^2\right)^2
 \right\}\right.\\\nonumber
&&\left.+ \sum_{i=1}^N\left\{2n_{f}\,
\left(1-\left(\frac{\alpha_{i}}{\pi}\Big|_{\rm mod\,2}-1\right)^2\right)^2
- 2\tilde n_{f}\,\left(1-\left(\left(\frac{\alpha_{i}}{\pi}+1\right)\Big|_{\rm mod \, 2}
-1\right)^2\right)^2
\right\}\right].
\eea
It is easy to see that $\alpha_i= 2\pi k/N$, $k=0,1,2,\ldots N-1$ are
extrema of the action. The contributions from the adjoint degrees of
freedom are naturally invariant under ${\mathbb Z}_N$ shifts
$\alpha_i\to \alpha_i + 2\pi k/N$, while the fundamental flavours break
the discrete symmetry. Expanding around each of the $N$ extrema, we
can show that they are each local {\em minima} labelled by the integer
$k$, provided
\be
N(2+ n_{A}+ \tilde n_{A})+ \tilde n_{f} 
\left(1-12\frac{k^2}{N^2}\right) + 2 n_{f} \,
\left(1- \frac{6k}{N}+ \frac{6k^2}{N^2}\right) > 0. 
\ee
As long as this condition is satisfied, at the 1-loop level, all
$N$ extrema of the effective potential will be local
minima. While it is interesting to ask what the fate of the effective
potential is, when the above condition is not satisfied, in the theory
that we are interested in (namely, ${\cal N}=4$ SYM coupled to $N_f$,
${\cal N}=2$ hypermultiplets) -- $n_{A}=6$, $\tilde n_{A}=4$,
$\tilde n_{f}=n_{f}=2
N_f$ and $N\gg N_f$ -- the extrema will always be local minima.

The global minimum of the effective potential is at $\alpha_i=0$.

The effect of the imaginary chemical potential for the fundamental
flavours is easily captured. Let us, for the sake of simplicity,
restrict attention to the case where the complex scalars and fermions in the
fundamental represention, all have the same (imaginary) chemical
potential. Then, an  imaginary chemical potential is introduced by the
replacement,
\be
\tilde D_0\to \partial_\tau+ i T {\cal A} - i \mu_I\,{\bf 1}.
\ee
Therefore, the effective potential for $\mu_I\neq 0$ is the same as 
(\ref{effpot}) with $\alpha_i \to \alpha_i - \beta \mu_I$. For small
non-zero  $\beta \mu_I$, the global minimum continues to be at
$\alpha_i=0$. However, when $\beta\mu_I = 2\pi r/N$, ($r\in {\mathbb
  Z}$) the global minimum shifts to 
$\alpha_i =2\pi r/N$. This is simply because the effective
potential from the fundamental flavours depends on
$\alpha_i-\beta\mu_I$, whilst the adjoint sector is independent of $\mu_I$.
One concludes that as $\beta\mu_I$ is increased from $0$, past $\pi/N$, the
global minimum jumps to $\alpha_i =2\pi/N$ following a first
order phase transition. The free energy of the theory as a function of
$\mu_I$ can now be calculated. This is done by first evaluating the 
the effective potential at a given 
global minimum $\alpha_i=0$,
\bea
&&{\cal F}(\beta\mu_I)\,\equiv \,V_{\rm eff} (\alpha_i=0)=\\\nonumber
&&N T^4\frac{\pi^2}{24}\left[
n_{f}\left(1-\left(-\frac{\beta\mu_I}{\pi}\Big|_{\rm mod \,
      2}-1\right)^2\right)^2
-\tilde n_{f}\left(1-\left(\left(-\frac{\beta\mu_I}{\pi}+1\right)\Big|_{\rm mod\, 2}-1\right)^2\right)^2
\right].
\eea
This yields the correct free energy, but only in the range
$-\frac{\pi}{N}< \beta\mu_I < \frac{\pi}{N}$. Outside this range, for
instance when $\beta\mu_I > \frac{\pi}{N}$, the global minimum moves
to $\alpha_i=2\pi/N$ and the free energy is given by the function
${\cal  F}(\beta\mu_I -2\pi/N)$. The correct free energy as a function
of $\mu_I$ is 
\be
F[\mu_I] = \min_{r\in {\mathbb Z}} \,\,{\cal F}\left(\beta\mu_I - 
\frac{2\pi r}{N}\right),
\ee
a function with cusps at $\beta\mu_I = (2r-1)\pi /N$, $r\in {\mathbb Z}$.

When $N$ is large, only the behaviour of $F$ near its local minima is
important, {\it i.e.}, 
\be
F[\mu_I]\Big|_{N\gg 1} = (2 n_{f}+\tilde n_{f})\,N\,\frac{T^4}{12}\,
\min_{r\in{\mathbb Z}}\,\left(\beta\mu_I-\frac{2\pi r}{N}\right)^2.
\ee
\\

{\bf \underline{Massive fundamental flavours:}} When $m\neq 0$, the
effective potential Eq. (\ref{effective}) cannot be expressed in
simple analytical form. Nevertheless, it has the same qualitative properties as
the $m_q=0$ potential in the presence of non-vanishing
$\mu_I$. Specifically, we can follow the reasoning above to deduce the
free energy at large $N$,
\be
F[\mu_I]= N\, \frac{T^4}{12} \,\left[n_{f} \,f_{\rm bose}\left(\frac{m_q}{T}\right)+ 
\tilde n_{f} \,f_{\rm fermi}\left(\frac{m_q}{T}\right)\right]\,
\min_{r\in {\mathbb Z}}\left(\beta\mu_I-\frac{2\pi r}{N}\right)^2,
\ee
where
\be
f_{\rm bose}\left(y\right)= \frac{3}{2\pi^2}\int_0^\infty x^2 \,dx\,
\sinh^{-2}\left(\frac{1}{2}\sqrt{x^2+y^2}\right),
\ee
and
\be
f_{\rm fermi}
\left(y\right)= \frac{3}{2\pi^2}\int_0^\infty x^2 \,dx\,
\cosh^{-2}\left(\frac{1}{2}\sqrt{x^2+y^2}\right).
\ee
As $m_q/T \to 0$, the functions $f_{\rm bose}$ and $f_{\rm fermi}$ 
approach $2$ and $1$
respectively.


\begin{thebibliography}{99}

\bibitem{maldacena}
  J.~M.~Maldacena,
  ``The large N limit of superconformal field theories and supergravity,''
  Adv.\ Theor.\ Math.\ Phys.\  {\bf 2}, 231 (1998)
  [Int.\ J.\ Theor.\ Phys.\  {\bf 38}, 1113 (1999)]
  [hep-th/9711200].

\bibitem{witten1}
  E.~Witten,
  ``Anti-de Sitter space, thermal phase transition, and confinement in  gauge
  theories,''
  Adv.\ Theor.\ Math.\ Phys.\  {\bf 2}, 505 (1998)
  [hep-th/9803131].

\bibitem{witten2}
  E.~Witten,
  ``Anti-de Sitter space and holography,''
  Adv.\ Theor.\ Math.\ Phys.\  {\bf 2}, 253 (1998)
  [hep-th/9802150].


\bibitem{gubser}
  S.~S.~Gubser,
  ``Thermodynamics of spinning D3-branes,''
  Nucl.\ Phys.\  B {\bf 551}, 667 (1999)
  [hep-th/9810225].

\bibitem{bcs}
  K.~Behrndt, M.~Cvetic and W.~A.~Sabra,
  ``Non-extreme black holes of five dimensional N = 2 AdS supergravity,''
  Nucl.\ Phys.\  B {\bf 553}, 317 (1999)
  [hep-th/9810227].

\bibitem{Chamblin:1999tk}
  A.~Chamblin, R.~Emparan, C.~V.~Johnson and R.~C.~Myers,
  ``Charged AdS black holes and catastrophic holography,''
  Phys.\ Rev.\  D {\bf 60}, 064018 (1999)
  [hep-th/9902170].


\bibitem{gc}
  M.~Cvetic and S.~S.~Gubser,
  ``Phases of R-charged black holes, spinning branes and strongly coupled
  gauge theories,''
  JHEP {\bf 9904}, 024 (1999)
  [hep-th/9902195]; 

\bibitem{Sundborg:1999ue}
  B.~Sundborg,
  ``The Hagedorn Transition, Deconfinement and N=4 SYM Theory,''
  Nucl.\ Phys.\  B {\bf 573}, 349 (2000)
  [hep-th/9908001].

\bibitem{Aharony:2003sx}
  O.~Aharony, J.~Marsano, S.~Minwalla, K.~Papadodimas and M.~Van Raamsdonk,
  ``The Hagedorn / deconfinement phase transition in weakly coupled large N
  gauge theories,''
  Adv.\ Theor.\ Math.\ Phys.\  {\bf 8}, 603 (2004)
  [hep-th/0310285].

\bibitem{Yamada:2006rx}
  D.~Yamada and L.~G.~Yaffe,
  ``Phase diagram of N = 4 super-Yang-Mills theory with R-symmetry chemical
  potentials,''
  JHEP {\bf 0609}, 027 (2006)
  [hep-th/0602074].


\bibitem{Yamada:2007gb}
  D.~Yamada,
  ``Metastability of R-charged black holes,''
  Class.\ Quant.\ Grav.\  {\bf 24}, 3347 (2007)
  [hep-th/0701254].


\bibitem{Hollowood:2008gp}
  T.~J.~Hollowood, S.~P.~Kumar, A.~Naqvi and P.~Wild,
  ``N=4 SYM on $S^3$ with Near Critical Chemical Potentials,''
  JHEP {\bf 0808}, 046 (2008)
  [0803.2822 [hep-th]].



\bibitem{karchkatz} 
 A.~Karch and L.~Randall,
  ``Open and closed string interpretation of SUSY CFT's on branes with
  boundaries,''
  JHEP {\bf 0106}, 063 (2001)
  [hep-th/0105132].

  A.~Karch and E.~Katz,
  ``Adding flavor to AdS/CFT,''
  JHEP {\bf 0206}, 043 (2002)
  [hep-th/0205236].





\bibitem{ss}
  T.~Sakai and S.~Sugimoto,
  ``Low energy hadron physics in holographic QCD,''
  Prog.\ Theor.\ Phys.\  {\bf 113}, 843 (2005)
  [hep-th/0412141].

\bibitem{babington}
  J.~Babington, J.~Erdmenger, N.~J.~Evans, Z.~Guralnik and I.~Kirsch,
  ``Chiral symmetry breaking and pions in non-supersymmetric gauge /  gravity
  duals,''
  Phys.\ Rev.\  D {\bf 69}, 066007 (2004)
  [hep-th/0306018].



\bibitem{kirsch}
  I.~Kirsch,
  ``Generalizations of the AdS/CFT correspondence,''
  Fortsch.\ Phys.\  {\bf 52}, 727 (2004)
  [hep-th/0406274].

\bibitem{ghoroku1}
  K.~Ghoroku, T.~Sakaguchi, N.~Uekusa and M.~Yahiro,
  ``Flavor quark at high temperature from a holographic model,''
  Phys.\ Rev.\  D {\bf 71}, 106002 (2005)
  [hep-th/0502088].


\bibitem{mateos1}
  D.~Mateos, R.~C.~Myers and R.~M.~Thomson,
  ``Holographic phase transitions with fundamental matter,''
  Phys.\ Rev.\ Lett.\  {\bf 97}, 091601 (2006)
  [hep-th/0605046].

\bibitem{albash}
  T.~Albash, V.~G.~Filev, C.~V.~Johnson and A.~Kundu,
  ``A topology-changing phase transition and the dynamics of flavour,''
  Phys.\ Rev.\  D {\bf 77}, 066004 (2008)
  [hep-th/0605088].

\bibitem{karch1}
  A.~Karch and A.~O'Bannon,
  ``Chiral transition of N = 4 super Yang-Mills with flavor on a 3-sphere,''
  Phys.\ Rev.\  D {\bf 74}, 085033 (2006)
  [hep-th/0605120].



\bibitem{mateos2}
  S.~Kobayashi, D.~Mateos, S.~Matsuura, R.~C.~Myers and R.~M.~Thomson,
  ``Holographic phase transitions at finite baryon density,''
  JHEP {\bf 0702}, 016 (2007)
  [hep-th/0611099].

\bibitem{ghoroku2}
  K.~Ghoroku, M.~Ishihara and A.~Nakamura,
  ``D3/D7 holographic Gauge theory and Chemical potential,''
  Phys.\ Rev.\  D {\bf 76}, 124006 (2007)
  [0708.3706 [hep-th]].

\bibitem{mateos3}
  D.~Mateos, R.~C.~Myers and R.~M.~Thomson,
  ``Thermodynamics of the brane,''
  JHEP {\bf 0705}, 067 (2007)
  [hep-th/0701132].

\bibitem{Nakamura:2007nx}
  S.~Nakamura, Y.~Seo, S.~J.~Sin and K.~P.~Yogendran,
  Prog.\ Theor.\ Phys.\  {\bf 120}, 51 (2008)
  [0708.2818 [hep-th]].

\bibitem{obannon}
  A.~Karch and A.~O'Bannon,
  ``Holographic Thermodynamics at Finite Baryon Density: Some Exact Results,''
  JHEP {\bf 0711}, 074 (2007)
  [0709.0570 [hep-th]].

\bibitem{mateos4}
  D.~Mateos, S.~Matsuura, R.~C.~Myers and R.~M.~Thomson,
  ``Holographic phase transitions at finite chemical potential,''
  JHEP {\bf 0711}, 085 (2007)
  [0709.1225 [hep-th]].

\bibitem{Aharony:2006da}
  O.~Aharony, J.~Sonnenschein and S.~Yankielowicz,
  Annals Phys.\  {\bf 322}, 1420 (2007)
  [arXiv:hep-th/0604161].


\bibitem{Horigome:2006xu}
  N.~Horigome and Y.~Tanii,
  JHEP {\bf 0701}, 072 (2007)
  [arXiv:hep-th/0608198].


\bibitem{rw}
  A.~Roberge and N.~Weiss,
``GAUGE THEORIES WITH IMAGINARY CHEMICAL POTENTIAL AND THE PHASES OF QCD,''
Nucl.\ Phys.\  B {\bf 275}, 734 (1986).

\bibitem{kruczenski}
  M.~Kruczenski, D.~Mateos, R.~C.~Myers and D.~J.~Winters,
  ``Meson spectroscopy in AdS/CFT with flavour,''
  JHEP {\bf 0307}, 049 (2003)
  [hep-th/0304032].



\bibitem{andythesis}
  A.~O'Bannon,
  ``Holographic Thermodynamics and Transport of Flavor Fields,''
  0808.1115 [hep-th].




\bibitem{DeTar:2009ef}
  C.~DeTar and U.~M.~Heller,
  ``QCD Thermodynamics from the Lattice,''
  Eur.\ Phys.\ J.\  A {\bf 41} (2009) 405
  [0905.2949 [hep-lat]].

\bibitem{deForcrand:2010ys}
  P.~de Forcrand,
  ``Simulating QCD at finite density,''
  PoS {\bf LAT2009} (2009) 010
  [1005.0539 [hep-lat]].



\bibitem{deForcrand:2002ci}
  P.~de Forcrand and O.~Philipsen,
  ``The QCD phase diagram for small densities from imaginary chemical
  potential,''
  Nucl.\ Phys.\  B {\bf 642} (2002) 290
  [hep-lat/0205016].

\bibitem{deForcrand:2003hx}
  P.~de Forcrand and O.~Philipsen,
  ``The QCD phase diagram for three degenerate flavors and small baryon
  density,''
  Nucl.\ Phys.\  B {\bf 673} (2003) 170
  [hep-lat/0307020].

\bibitem{deForcrand:2006pv}
  P.~de Forcrand and O.~Philipsen,
  ``The chiral critical line of N(f) = 2+1 QCD at zero and non-zero baryon
  density,''
  JHEP {\bf 0701} (2007) 077
  [hep-lat/0607017].

\bibitem{deForcrand:2010he}
  P.~de Forcrand and O.~Philipsen,
  ``Constraining the QCD phase diagram by tricritical lines at imaginary
  chemical potential,''
  1004.3144 [hep-lat].

\bibitem{D'Elia:2002gd}
  M.~D'Elia and M.~P.~Lombardo,
  ``Finite density QCD via imaginary chemical potential,''
  Phys.\ Rev.\  D {\bf 67} (2003) 014505
  [hep-lat/0209146].

\bibitem{D'Elia:2004at}
  M.~D'Elia and M.~P.~Lombardo,
  ``QCD thermodynamics from an imaginary mu(B): Results on the four flavor
  lattice model,''
  Phys.\ Rev.\  D {\bf 70} (2004) 074509
  [hep-lat/0406012].

\bibitem{D'Elia:2007ke}
  M.~D'Elia, F.~Di Renzo and M.~P.~Lombardo,
  ``The strongly interacting Quark Gluon Plasma, and the critical behaviour
  of QCD at imaginary chemical potential,''
  Phys.\ Rev.\  D {\bf 76} (2007) 114509
  [0705.3814 [hep-lat]].

\bibitem{D'Elia:2009tm}
  M.~D'Elia and F.~Sanfilippo,
  ``Thermodynamics of two flavor QCD from imaginary chemical potentials,''
  Phys.\ Rev.\  D {\bf 80} (2009) 014502
  [0904.1400 [hep-lat]].

\bibitem{D'Elia:2009qz}
  M.~D'Elia and F.~Sanfilippo,
  ``The order of the Roberge-Weiss endpoint (finite size transition) in QCD,''
  Phys.\ Rev.\  D {\bf 80} (2009) 111501
  [0909.0254 [hep-lat]].
                                                                                
\bibitem{Cea:2010md}
  P.~Cea, L.~Cosmai, M.~D'Elia and A.~Papa,
  ``The phase diagram of QCD with four degenerate quarks,''
  1004.0184 [hep-lat].







\bibitem{Braun:2009gm}
   J.~Braun, L.~M.~Haas, F.~Marhauser and J.~M.~Pawlowski,
   arXiv:0908.0008 [hep-ph].


\bibitem{Bluhm:2007cp}
  M.~Bluhm and B.~Kampfer,
  ``Quasiparticle Model of Quark-Gluon Plasma at Imaginary Chemical
  Potential,''
  Phys.\ Rev.\  D {\bf 77} (2008) 034004
  [0711.0590 [hep-ph]].


\bibitem{Sakai:2008um}
  Y.~Sakai, K.~Kashiwa, H.~Kouno and M.~Yahiro,
  ``Phase diagram in the imaginary chemical potential region and extended Z3
  symmetry,''
  Phys.\ Rev.\  D {\bf 78} (2008) 036001
  [0803.1902 [hep-ph]].

\bibitem{Kashiwa:2008bq}
  K.~Kashiwa, M.~Matsuzaki, H.~Kouno, Y.~Sakai and M.~Yahiro,
  ``Meson mass at real and imaginary chemical potentials,''
  Phys.\ Rev.\  D {\bf 79} (2009) 076008
  [0812.4747 [hep-ph]].

\bibitem{Kouno:2009bm}
  H.~Kouno, Y.~Sakai, K.~Kashiwa and M.~Yahiro,
  ``Roberge-Weiss phase transition and its endpoint,''
  J.\ Phys.\ G {\bf 36} (2009) 115010
  [0904.0925 [hep-ph]].




\bibitem{weiss}
  N.~Weiss,
  ``The Effective Potential For The Order Parameter Of Gauge Theories At Finite
  Phys.\ Rev.\  D {\bf 24}, 475 (1981).

  N.~Weiss,
  ``The Wilson Line In Finite Temperature Gauge Theories,''
  Phys.\ Rev.\  D {\bf 25}, 2667 (1982).




\bibitem{FG}
C. Fefferman and C. Robin Graham, 'Conformal Invariants,' in {\it
  Elie Cartan et les Mathematiques d'aurjourd'hui} (Asterique, 1985) 95.

\bibitem{karchskenderis}
  A.~Karch, A.~O'Bannon and K.~Skenderis,
  ``Holographic renormalization of probe D-branes in AdS/CFT,''
  JHEP {\bf 0604}, 015 (2006)
  [hep-th/0512125].


\bibitem{aharonywitten}
  O.~Aharony and E.~Witten,
  ``Anti-de Sitter space and the center of the gauge group,''
  JHEP {\bf 9811}, 018 (1998)
  [hep-th/9807205].

\bibitem{malpol}
  J.~M.~Maldacena,
  ``Wilson loops in large N field theories,''
  Phys.\ Rev.\ Lett.\  {\bf 80}, 4859 (1998)
  [hep-th/9803002].

  S.~J.~Rey and J.~T.~Yee,
  ``Macroscopic strings as heavy quarks in large N gauge theory and  anti-de
  Sitter supergravity,''
  Eur.\ Phys.\ J.\  C {\bf 22}, 379 (2001)
  [hep-th/9803001].


\bibitem{yee}
  H.~U.~Yee,
  ``Fate of Z(N) domain wall in hot holographic QCD,''
  JHEP {\bf 0904}, 029 (2009)
  [0901.0705 [hep-th]].


\bibitem{mst}
  R.~C.~Myers, A.~O.~Starinets and R.~M.~Thomson,
  ``Holographic spectral functions and diffusion constants for fundamental
  matter,''
  JHEP {\bf 0711}, 091 (2007)
  [0706.0162 [hep-th]].

\bibitem{hoyos}
  C.~Hoyos-Badajoz, K.~Landsteiner and S.~Montero,
  ``Holographic Meson Melting,''
  JHEP {\bf 0704}, 031 (2007)
  [hep-th/0612169].


\bibitem{liu}
  T.~Faulkner and H.~Liu,
  ``Condensed matter physics of a strongly coupled gauge theory with quarks:
  some novel features of the phase diagram,''
  0812.4278 [hep-th].
  T.~Faulkner and H.~Liu,
  ``Meson widths from string worldsheet instantons,''
  Phys.\ Lett.\  B {\bf 673}, 161 (2009)
  [0807.0063 [hep-th]].




\bibitem{benincasa}
  P.~Benincasa,
  0911.0075 [Unknown].



\bibitem{sbh}
  T.~Hollowood, S.~P.~Kumar and A.~Naqvi,
  ``Instabilities of the small black hole: A view from N = 4 SYM,''
  JHEP {\bf 0701}, 001 (2007)
  [hep-th/0607111].


\end{thebibliography}
\end{document}